\documentclass[aps,prl,reprint,superscriptaddress]{revtex4-1}

\usepackage{graphicx}
\usepackage{dcolumn}
\usepackage{bm}
\usepackage{color}
\usepackage{xcolor}
\usepackage{epstopdf}
\usepackage{gensymb}
\usepackage{ulem}
\usepackage{float}
\usepackage{xr}
\usepackage{sidecap}
\usepackage{mathrsfs}
\usepackage{amsmath}
\usepackage{amsthm}
\usepackage{enumerate}
\usepackage{soul}
\usepackage{amssymb}
\usepackage{units}

\begin{document}

\preprint{APS/123-QED}

\title{Rolling Waves with Non-Paraxial Phonon Spins\\}





\author{Peng Zhang}
\affiliation{Department of Mechanical Engineering, University of Utah, Salt Lake City, UT 84112, USA}

\author{Christian Kern}
\affiliation{Department of Mathematics, University of Utah, Salt Lake City, UT 84112, USA}

\author{Sijie Sun}
\affiliation{Harvard John A. Paulson School of Engineering and Applied Science, Harvard University, Cambridge, MA 02138, USA}

\author{David A. Weitz}
\affiliation{Harvard John A. Paulson School of Engineering and Applied Science, Harvard University, Cambridge, MA 02138, USA}

\author{Pai Wang}%
\affiliation{Department of Mechanical Engineering, University of Utah, Salt Lake City, UT 84112, USA}
\thanks{pai.wang@utah.edu}

\date{Submitted for review on July 02, 2020}

\begin{abstract}
We demonstrate a new class of elastic waves in the bulk: When longitudinal and transverse components propagate at the same speed, rolling waves with a spin that is not parallel to the wave vector can emerge. First, we give a general definition of spin for traveling waves. Then, since rolling waves cannot exist in isotropic solids, we derive conditions for anisotropic media and proceed to design architected materials capable of hosting rolling waves. Numerically, we show spin manipulations by reflection. Structures reported in this work can be fabricated using available techniques, opening new possibilities for spin technologies in acoustics, mechanics and phononics.
\end{abstract}
\maketitle
The intrinsic spin angular momentum is an important property not only in quantum mechanical descriptions of fundamental particles but also in polarization representations of general wave mechanics~\cite{long2018,burns2020acoustic,shi2019observation,burns2020acoustic}. For electromagnetic and elastic waves, the transversely circular polarization has a direct correspondence to the spin-1 photons~\cite{banzer2012,eismann2020} and phonons~\cite{zhu2018observation,holanda2018detecting,shi2020,ruckriegel2020angular,an2020coherent}, respectively. This duality between the classical and quantum worlds has inspired a number of recent studies in optics~\cite{le2015,sollner2015,peng2019transverse,kim2012time}, gravitational waves~\cite{golat2020evanescent}, acoustics~\cite{wang2018topological,toftul2019acoustic,bliokh2019spin,bliokh2019acoustic,long2020symmetry} and solid mechanics~\cite{calderin2019,kumar2020,hasan2020experimental}. Many intriguing features have been demonstrated, including spin-orbit coupling~\cite{liu2017circular,fu2019}, spin-momentum locking~\cite{liu2019three} and topological edge states analogous to the quantum spin Hall effect~\cite{zhou2018quantum,wu2018topological,tuo2019twist,jin2020}.\\
\begin{figure}[b!]
\includegraphics[scale=0.3]{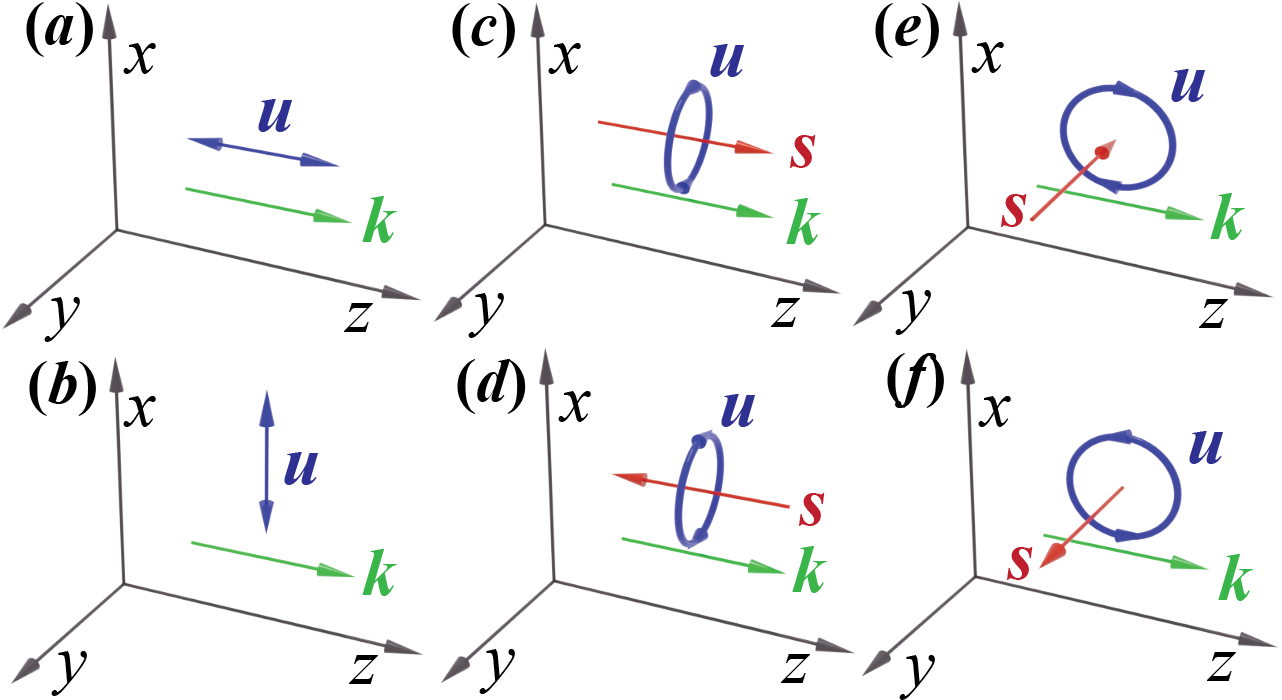}
\caption{\label{fig:F1} Spin categories of bulk elastic waves: (a) linear longitudinal $(m,n,l)=(0,0,1)$ and (b) linear transverse $(1,0,0)$ are spin-less. (c) right circular $(1,i,0)$ and (d) left circular $(1,-i,0)$ carry paraxial spins. (e) rolling forward $(1,0,i)$ and (f) rolling backward $(1,0,-i)$ carry non-paraxial spins. Here $\boldsymbol{u}$ shows displacement trajectory. $\boldsymbol{k}$ and $\boldsymbol{s}$ represent wave vector and spin vector, respectively. }
\end{figure}
Focusing on elastic waves in solids, we note that there is naturally a \textit{traveling} longitudinal component (Fig.\,\ref{fig:F1}(a)), which can co-propagate with shear waves (Fig.\,\ref{fig:F1}(b)). This is in stark contrast with electromagnetic waves, for which the longitudinal component can only exist either due to localized interference~\cite{banzer2012,eismann2020}, with field couplings~\cite{sollner2015,le2015,peng2019transverse}, or as evanescent waves~\cite{kim2012time}. Consequently, in addition to the usual paraxial spin carried by shear waves (Figs.\,\ref{fig:F1}(c) and \ref{fig:F1}(d)), elastic waves may also carry spins corresponding to the displacement trajectory rolling forward (Fig.\,\ref{fig:F1}(e)) or rolling backward (Fig.\,\ref{fig:F1}(f)). Such special cases of non-paraxial spins are also referred to as ``transverse spins"~\cite{leykam2020edge,wei2020far}, as the spin vector here is perpendicular to the wave vector. Importantly, recent research has proposed hybrid spins of two elastic waves using interference patterns, i.e., a localized superposition of waves in different directions~\cite{long2018}. Differently, here we report new results on \textit{traveling} rolling waves with \textit{propagating} non-paraxial spins, which are defined as follows: We consider the displacement field of a general plane wave $\boldsymbol{u}=\boldsymbol{\tilde{u}}\exp(i\boldsymbol{k \cdot r}-i\omega t)$ with 
\begin{equation}
     \boldsymbol{\tilde{u}} =\frac{A}{\sqrt{|m|^2+|n|^2+|l|^2}} \left( \begin{array}{c}
      m  \\ n \\  l 
    \end{array}\right),
\label{u_hat}
\end{equation}
where $A$ denotes the amplitude. Importantly, here $m$, $n$ and $l$ are complex-valued, so that they contain the information about not only relative amplitudes but also \textit{relative phase differences} among the displacement components.
The spin angular momentum density, as a real-valued vector, can be calculated as~\cite{long2018,burns2020acoustic,berry2009optical}:
\begin{equation}
    \boldsymbol{s}= \frac{\rho \omega}{2} \textless\boldsymbol{\tilde{u}}|\boldsymbol{\hat{\textbf{S}}}|\boldsymbol{\tilde{u}}\textgreater = \frac{\rho \omega}{2} {\rm Im}[\boldsymbol{\tilde{u}}^*\times \boldsymbol{\tilde{u}}],
\label{S1}\end{equation}
where $(\cdot)^*$ denotes complex conjugation, and the spin-1 operator is defined as
\begin{equation}
    \boldsymbol{\hat{\textbf{S}}}=\left[\left( \begin{array}{ccc}
        0 & 0 & 0\\
        0 & 0 & -i\\
        0 & i & 0
    \end{array} \right),
    \left( \begin{array}{ccc}
        0 & 0 & i\\
        0 & 0 & 0\\
        -i & 0 & 0
    \end{array} \right),
    \left( \begin{array}{ccc}
        0 & -i & 0\\
        i & 0 & 0\\
        0 & 0 & 0
    \end{array} \right) \right]. 
\end{equation}
Hence, the spin density for a general traveling wave is  
\begin{equation}
    \boldsymbol{s}=\frac{\rho\omega A^2}{|m|^2+|n|^2+|l|^2}{\rm Im}\left( \begin{array}{c}
      n^*l  \\ l^*m \\  m^*n 
    \end{array}\right).
\label{S2}\end{equation}
For stable propagation of rolling polarizations (Figs.\,\ref{fig:F1}(e) or \ref{fig:F1}(f)), the longitudinal and transverse components have to share the same phase velocity $c = \omega / k$. 
While this condition may be satisfied due to the effects of boundaries and interfaces, e.g., acoustic waves in air ducts~\cite{long2020symmetry}, water waves in the ocean~\cite{li2018}, Rayleigh waves along solid surfaces~\cite{brule2020possibility,zhao2020non} and Lamb waves in elastic plates~\cite{jin2020}, we can show that it is never satisfied for bulk waves in isotropic media.\\
Consider an isotropic elastic material with shear modulus $\mu$, Poisson’s ratio $\nu$, and mass density $\rho$: The ratio between the transverse phase velocity, $c_\text{T}$, and the longitudinal phase velocity, $c_\text{L}$, is given by
\begin{equation}
    \frac{c_\text{T}}{c_\text{L}} = 
    {\sqrt{\frac{\mu}{\rho}}} \Bigg/ {\sqrt{\frac{2\mu(1-\nu)}{(1-2\nu)\rho}}} =\sqrt{1-\frac{1}{2(1-\nu)}}.
\label{W1}
\end{equation}
For static stability~\cite{Landau1970}, we are constrained by $-1\leq \nu \leq 1/2 $ in 3D and $-1\leq \nu \leq 1$ in 2D, both of which imply ${c_\text{T}}/{c_\text{L}} \leq \sqrt{3}/2$. Even if we disregard these constraints and allow for arbitrary values of the Poisson’s ratio, the speed ratio ${c_\text{T}}/{c_\text{L}}$ can only asymptotically approach unity when $\nu \rightarrow \infty$. Therefore, as a frequency-independent material property, the \textit{equal-speed criterion}, ${c_\text{T}}={c_\text{L}}$, cannot be met by any isotropic medium.\\
As a consequence, we turn our attention to anisotropic media. For a plane wave with wave vector $\bm{k}$ and wave number $k=|\bm{k}|$, we write
\begin{equation}
    \boldsymbol{\tilde{k}}=\bm{k}/k=l_1\boldsymbol{e}_1+l_2\boldsymbol{e}_2+l_3\boldsymbol{e}_3,
\label{eq:72}\end{equation}
where $l_1,l_2,l_3$ are the direction cosines of the wave vector with respect to the Cartesian coordinate axes. We define a matrix $\textbf{L}$ as
\begin{equation}
    \textbf{L}=\left( \begin{array}{cccccc}
    l_1 & 0 & 0 & 0 & l_3 & l_2\\
    0 & l_2 & 0 & l_3 & 0 & l_1\\
    0 & 0 & l_3 & l_2 & l_1 & 0
\end{array} 
\right ),
\end{equation}
and introduce the Kelvin-Christoffel matrix~\cite{carcione2007wave,SI}:
\begin{equation}
    \boldsymbol{\Gamma}=\textbf{L} \boldsymbol{\cdot} \textbf{C} \boldsymbol{\cdot} \textbf{L}^\text{T}\quad \text{or} \quad \mathit{\Gamma}_{ij}=L_{iI}C_{IJ}L_{Jj},
\end{equation}
where $C_{IJ}$ is the elastic stiffness in Voigt notation $(I,J=1,2,3...6)$.
Then, with the definition of phase velocity $c=\omega / k$, the wave equation can be written as~\cite{carcione2007wave,SI}:
\begin{equation}
    \boldsymbol{\Gamma \cdot u}-\rho c^2 \boldsymbol{u}=(\boldsymbol{\Gamma}-\rho c^2\textbf{I}) \cdot\boldsymbol{u}=\boldsymbol{0}.
\label{eq:75}\end{equation}
Therefore, the \textit{equal-speed criterion} is equivalent to the 3-by-3 matrix $\boldsymbol{\Gamma}$ having degenerate eigenvalues. For media with spatial symmetries, the criterion can be simplified further. Here we consider three examples for bulk waves propagating along the $z$-direction:
\begin{subequations}
\label{criterion}
\begin{align}
    &\text{2D $xz$-plane strain: } C_{33} = C_{55} \text{ and } C_{35} = 0\label{criterion-2d}\\ 
    &\text{Transversely $xy$-isotropic: } C_{33} = C_{44}\label{criterion-trans}\\
    &\text{Cubic symmetric: } C_{11} = C_{44}\label{criterion-cubic}
\end{align}
\end{subequations}
To the best of our knowledge, the criteria listed in Eq.\,(\ref{criterion}) are not satisfied by any existing materials, natural or synthetic. This motivates us to design metamaterials for this purpose. Aiming for structures that can be readily fabricated, we exclusively focus on architected geometries made from a single isotropic base material with Young's modulus $E$ and Poisson's ratio $\nu = 0.3$. To identify suitable designs that work in the long wavelength limit, we perform quasi-static calculations using unit cells with appropriate periodic boundary conditions~\cite{overvelde2014relating} on the finite element platform \textsc{abaqus} (element types \textsc{cpe4} and \textsc{c3d4}). From the numerical results, we extract the dimensionless effective elastic constants $\bar C_{IJ}=C_{IJ}/E$.\\
First, we consider the 2D $xz$-plane strain case for waves propagating along the $z$-direction. In order to arrive at a micro-structure satisfying the equal-speed criterion, we need to strengthen the shear stiffness of the material without increasing its normal stiffness. Fig.\,2(a) shows an example  unit-cell, where the X-shaped center enhances the shear stiffness and the absence of vertical support reduces the normal stiffness. Using the numerical procedure described above, we calculate the dependence of the dimensionless effective elastic constants $\bar{C}_{33}$ and $\bar{C}_{55}$ on the geometry parameters $L_1$ and $L_2$. The results are presented as two surfaces shown in Fig.\,2(b). The equal-speed criterion is met at the line of intersection of the two surfaces. The geometry shown in Fig.\,2(a) corresponds to the circled point at $(L_1/a,L_2/a)=(0.2,0.1962)$.
Next, as shown in Fig.\,2(c), we present another 2D micro-structure satisfying Eq.\,(\ref{criterion}a). This design was adapted from a previous study on dilatational materials~\cite{buckmann2014three}, which in turn was based on earlier theoretical results~\cite{milton2013complete}. The unit cell entails $C_{11} = C_{33}$ due to symmetry, and, hence, supports the propagation of rolling waves along both the $x$- and $z$-direction. The circled point $(b_1/a,b_2/a)=(0.01878,0.005)$ in Fig.\,2(d) corresponds to the geometry in Fig.\,2(c) with $b_3=b_4=0.05a$ and $b_5=0.3221a$. While this geometry was previously designed for auxetic (i.e., negative Poisson's ratio) properties~\cite{buckmann2014three}, the parameters satisfying the equal-speed criterion actually result in a positive effective 2D Poisson's ratio between the principal directions, $\nu_{xz} = 0.996$. As this structure is strongly anisotropic, the fact that the principal Poisson's ratio approaches unity does not imply a large difference between $c_\text{T}$ and $c_\text{L}$.
We note that both 2D designs presented in Fig.\,2 are mirror-symmetric with respect to both the $x$- and $z$-axis. This symmetry implies the absence of normal-to-shear or shear-to-normal couplings in the effective constitutive relations. Thus, irregardless of the geometry parameters, the condition $C_{35} = 0$ in Eq.\,(\ref{criterion}a),  holds for both structures."\\
\begin{figure}[htb]
\centering
\includegraphics[scale=0.36]{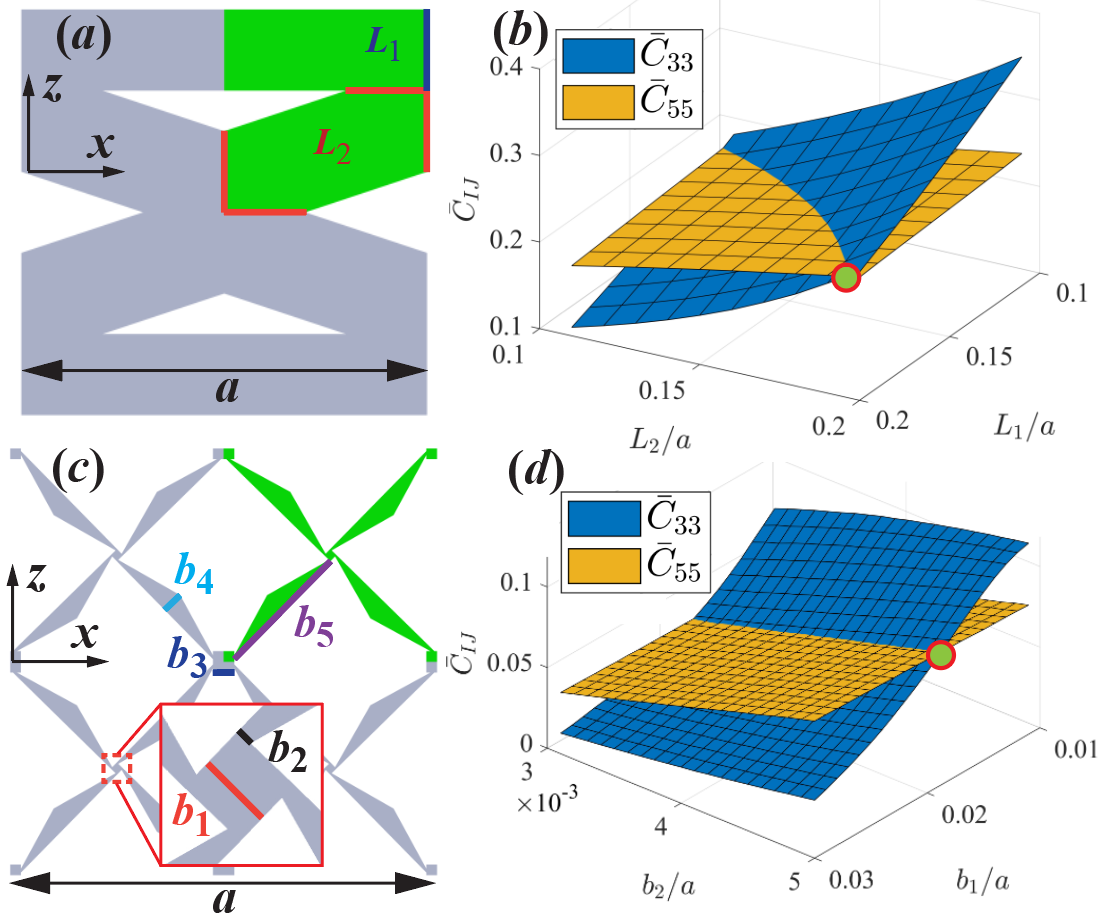}
\caption{\label{fig:F2} 2D metamaterials capable of hosting rolling waves: (a) Unit cell design obtained by taking mirror images of the green quarter. Each red straight line segment is of length $L_2$. (b) Numerically calculated effective elastic constants for the unit cell in (a) with varying geometric parameters. (c) Unit cell design adapted from \cite{buckmann2014three}. It is obtained by taking mirror images of the green quarter. (d) Numerical results for the unit cell in (c) with $b_3=b_4=0.05a$ and $b_5=0.3221a$. Geometries in (a) and (c) correspond to the circled points in (b) and (d), respectively.
}
\end{figure}
In the 3D case, we consider highly symmetric geometries exhibiting either transverse isotropy or cubic symmetry. Fig.\,3(a) shows the unit-cell design satisfying Eq.\,(\ref{criterion-trans}). The honeycomb geometry is chosen to guarantee isotropy in the $xy$-plane. Each out-of-plane wall is constructed by extruding the planar pattern in Fig.\,2(a). Besides $L_1$ and $L_2$, the wall thickness, $h$, is an additional parameter of the 3D structure. Numerical results for the dimensionless constants, $\bar C_{33}$ and $\bar C_{44}$ are shown in Fig.\,3(b) with $h/a=0.2$ fixed. At the line of surface intersection, we obtain a set of designs satisfying the equal-speed criterion. The geometry in Fig.\,3(a) corresponds to the circled point $(L_1/a,L_2/a)=(0.216,0.1)$ in Fig.\,3(b).\\
For the cubic symmetric case, a unit cell satisfying Eq.\,(\ref{criterion-cubic}) is shown in Fig.\,3(c). This geometry was previously studied as an auxetic micro-structure~\cite{dirrenberger2013effective}. It has the symmetry of the point group m$\bar{3}$m - one of the cubic crystallographic point groups, which are characterized by four axes of three-fold rotational symmetry. The symmetry axes can be identified with the body diagonals of the cubic unit cell~\cite{authier2003international,buckmann2014three,dirrenberger2013effective}. The cubic symmetry guarantees that the resulting stiffness matrix has a specific form with three independent elastic constants~\cite{authier2003international,norris2006poisson,SI}. The unit cell in Fig.\,3(c) is composed of identical beams with a square cross-section with side length $L$. For each of the six cubic faces, four beams extend from the vertices and meet at an interior point at a distance $h_2$ from the planar face center. By varying both $L$ and $h_2$, as shown in Fig.\,3(d), we numerically find the line of intersection where the equal-speed criterion is satisfied. Parameters at the circled point $(L/a,h_2/a)=(0.06,0.1035)$ correspond to the geometry in Fig.\,3(c). While being auxetic in other directions, this structure actually has a positive effective Poisson's ratio along its principal directions, as also shown in \cite{dirrenberger2013effective}.\\
\begin{figure}[t]
\centering
\includegraphics[scale=0.36]{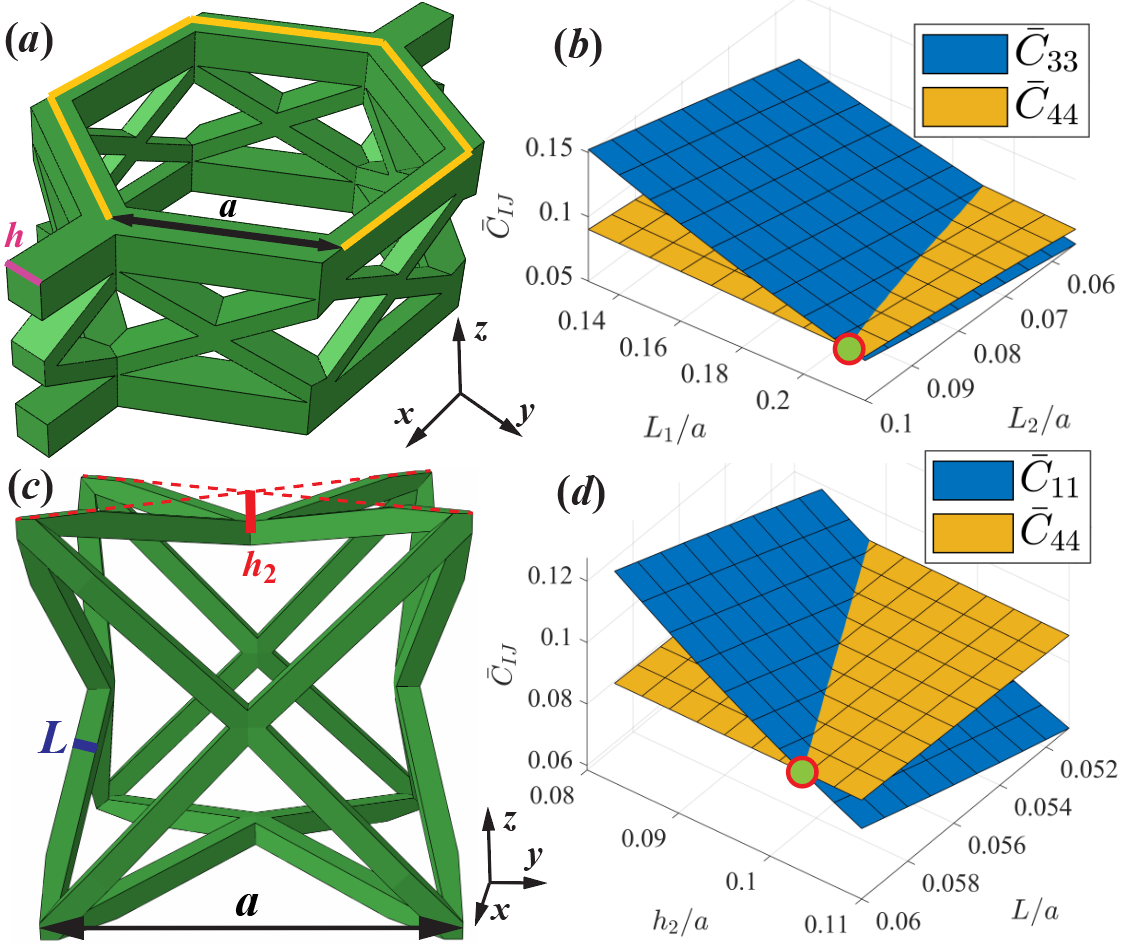}
\caption{\label{fig:F2} 
3D metamaterials capable of hosting rolling waves: (a) Honeycomb unit cell design based on the planar pattern shown in Fig.\,2(a), exhibiting isotropy in the $xy$-plane. (b) Numerical results of effective elastic constants for the unit cell in (a) with $h=0.2a$. (c) Unit cell design adapted from \cite{dirrenberger2013effective}, exhibiting cubic symmetry.  (d) Numerical results of the cubic case. Geometries in (a) and (c) correspond to the circled points in (b) and (d), respectively.
}
\end{figure}
As an example of non-paraxial spin manipulation, we next investigate normal reflections of a rolling wave at a general elastic boundary. Considering a rolling wave along the $z$-direction normally incident on a flat surface, we have the instantaneous wave displacement fields at time $t=0$ as $\boldsymbol{u}^\text{I}=\boldsymbol{\tilde{u}}^\text{I}\exp(\text{i}kz)$ and $\boldsymbol{u}^\text{R}=\boldsymbol{\tilde{u}}^\text{R}\exp(-\text{i}kz)$ with
\begin{equation}
    \boldsymbol{\tilde{u}}^\text{I}=\left( \begin{array}{c}
      m^\text{I}  \\ n^\text{I} \\ l^\text{I}
    \end{array}\right) \quad \text{and} \quad 
    \boldsymbol{\tilde{u}}^\text{R}=\left( \begin{array}{c}
      m^\text{R}  \\ n^\text{R} \\ l^\text{R}
    \end{array}\right),
\end{equation}
where the superscripts, I and R, denote the incident and reflected waves, respectively. For the surface at $z=0$, we have
\begin{equation}
    \sigma^0_{zj}=K_ju^0_j,\ \ j=x,y,z,
\label{R1}\end{equation}
where $K_j$ represents the distributed stiffness of a elastic foundation. At the surface, the stress $\sigma^0_{zj}$ and displacement $u^0_j$ are the superposition of the incident and reflected waves. By calculating the stresses and substituting them into the boundary conditions of \eqref{R1}, we obtain
\begin{equation}
    m^\text{R}=R_x m^\text{I},\ \ n^\text{R}=R_y n^\text{I},\ \ l^\text{R}=R_z l^\text{I},
\end{equation}
with 
\begin{equation}
    R_j=\frac{C_{33} ik-K_j}{C_{33} ik+K_j},\ \ j=x,y,z. 
\end{equation}
\begin{figure}[t]
\centering
\includegraphics[scale=0.55]{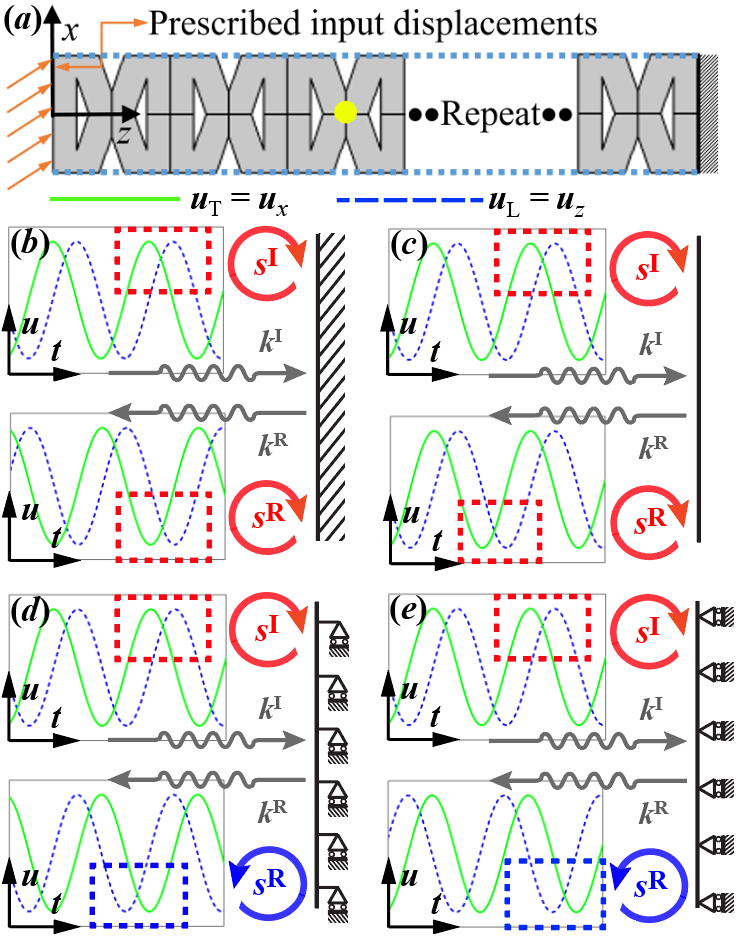}
\caption{\label{fig:F4} Reflections of a rolling wave (with $s^\text{I}_y=-1$) for normal incidence: (a) Illustration of the setup for the time-domain numerical simulations. Periodic boundary conditions are applied on the top and bottom edges (dotted lines). A total number of 70 unit cells are used along the $z$-direction. Insets of (b)-(e) show the time history of displacements at the position marked by the yellow dot (\textcolor{yellow}{$\bullet$}) in (a). 
(b) rigid and (c) free boundaries are both spin-preserving. (d) free-fixed and (e) fixed-free boundaries are both spin-flipping.}
\end{figure}
We next focus on 2D rolling waves in the $xz$-plane where the wave amplitudes in the $y$-direction vanish, $n^\text{I}=n^\text{R}=0$. For $K_x=K_z=0$, the boundary becomes traction-free (Neumann type) and we have $\boldsymbol{\tilde{u}}^\text{R}=\boldsymbol{\tilde{u}}^\text{I}$ with no phase change. For $K_x=K_z=\infty$, the boundary becomes rigid (Dirichlet type) and we have an out-of-phase reflected wave with $\boldsymbol{\tilde{u}}^\text{R}=-\boldsymbol{\tilde{u}}^\text{I}$. In both cases, we can obtain from Eq.\,(\ref{S1}) that $\boldsymbol{s}^\text{R}=\boldsymbol{s}^\text{I}$, so the spin is preserved. In contrast, for hybrid boundaries ($K_x=0$, $K_z=\infty$) and ($K_x=\infty$, $K_z=0$), similar analyses~\cite{SI} result in $\boldsymbol{s}^\text{R}=-\boldsymbol{s}^\text{I}$, so the spin is flipped due to the difference in phase change between the longitudinal and transverse components during the reflection process. These behaviors are further demonstrated in time-domain simulations of a metamaterial made of unit cells shown in Fig.\,2(a) using the commercial software \textsc{comsol} (quadratic quadrilateral elements). Fig.\,4 shows the results for the incident rolling wave carrying a non-paraxial spin of $s_y^\text{I}=-1$. The reflection process is spin-preserving in both rigid and free boundaries, while being spin-flipping for both hybrid free-fixed and hybrid fixed-free boundary conditions. Detailed numerical procedures and additional results of time-domain simulations are available in the Supplemental Material~\cite{SI}.\\
In summary, we studied elastic waves carrying non-paraxial spins, which can propagate in special anisotropic media satisfying the \textit{equal-speed criterion}, $c_\text{T}=c_\text{L}$. We presented both 2D and 3D metamaterial designs satisfying this criterion. In addition, we analysed the reflection of such rolling waves incident on elastic boundaries, demonstrating spin-preserving and spin-flipping behaviors. In contrast to scattering-based~\cite{he2018topological,deng2018metamaterials,celli2019bandgap,wang2020evanescent,xia2020experimental,rosa2020topological,ramakrishnan2020multistable} and resonance-based~\cite{palermo2019tuning,arretche2019experimental,wang2020robust,sugino2020,hussein2020thermal,ghanem2020nanocontact,nassar2020polar,bilal2020enhancement} metamaterials, our designs work in the  non-resonant long-wavelength regime~\cite{zheng2019theory,patil20193d,behrou2020topology,zheng2020non,xu2020physical}, essentially using exotic quasi-static properties for wave manipulations. All features shown in this study are \textit {frequency independent} up to the cutoff threshold, which is only limited by how small we can make the unit cells. The tailored structures can be readily fabricated by existing techniques~\cite{liu2020maximizing,elder2020nanomaterial}. This work lays a solid foundation for the new field of broadband phononic spin engineering.\\

This work was supported by start-up funds of the Dept. Mechanical Engineering at Univ.\,Utah. CK is supported by the NSF through Grant DMS-1814854. The authors thank Graeme Milton (Univ.\,Utah), Stephano Gonella (Univ.\,Minnesota), Liyuan Chen (Harvard Univ. \& Westlake Univ.) and Bolei Deng (Harvard Univ.) for discussions. The support and resources from the Center for High Performance Computing at Univ. Utah are gratefully acknowledged.

\normalem
%

\bibliography{ref_NoTitle}      

\end{document}


\title{Supporting Information for \\
\emph{Rolling Waves with Non-Paraxial Phonon Spins}}

\maketitle

\author{Peng Zhang}
\affiliation{Department of Mechanical Engineering, University of Utah, Salt Lake City, UT 84112, USA}

\author{Christian Kern}
\affiliation{Department of Mathematics, University of Utah, Salt Lake City, UT 84112, USA}

\author{Sijie Sun}
\affiliation{Harvard John A. Paulson School of Engineering and Applied Science, Harvard University, Cambridge, MA 02138, USA}

\author{David A. Weitz}
\affiliation{Harvard John A. Paulson School of Engineering and Applied Science, Harvard University, Cambridge, MA 02138, USA}

\author{Pai Wang}%
\affiliation{Department of Mechanical Engineering, University of Utah, Salt Lake City, UT 84112, USA}
\thanks{pai.wang@utah.edu}




\clearpage
\tableofcontents

\clearpage
\section{Other non-paraxial spins}
There is no limit to the directions of spin vectors of elastic waves, or equivalently, phonon spins in general. In addition to the examples shown in the main text, Fig.\,S1 shows more categories of traveling waves with various types of propagating non-paraxial spins. This full range of spin degrees of freedom brings great potential in future applications.\\

\begin{figure}[htb]
\centering
\includegraphics[width=0.5\textwidth]{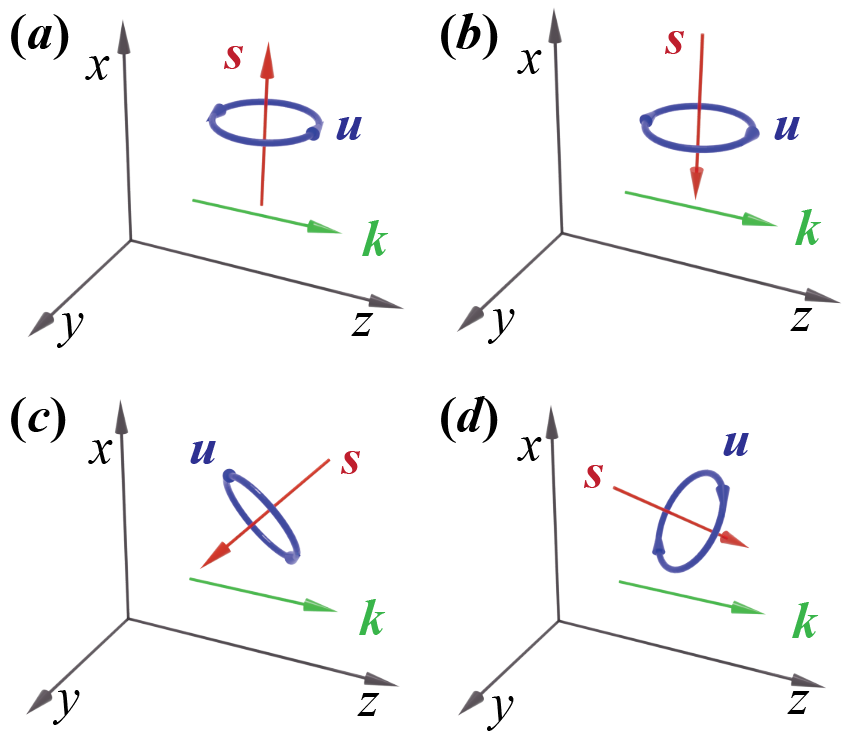}
\caption{\label{fig:F1}  The extended spin categories: (a,b) Spin vector perpendicular to wavenumber vector with $(m,n,l) = (i,0,1)$ and $(-i,0,1)$, respectively.  (c,d) Other non-paraxial spin directions with $(m,n,l) = (1+i,2,1-i)$ and $(1-i,1+i,2)$, respectively. }
\end{figure}

\clearpage
\section{Wave speeds in isotropic materials}

\begin{figure}[htb]
\centering
\includegraphics[width=0.7\textwidth]{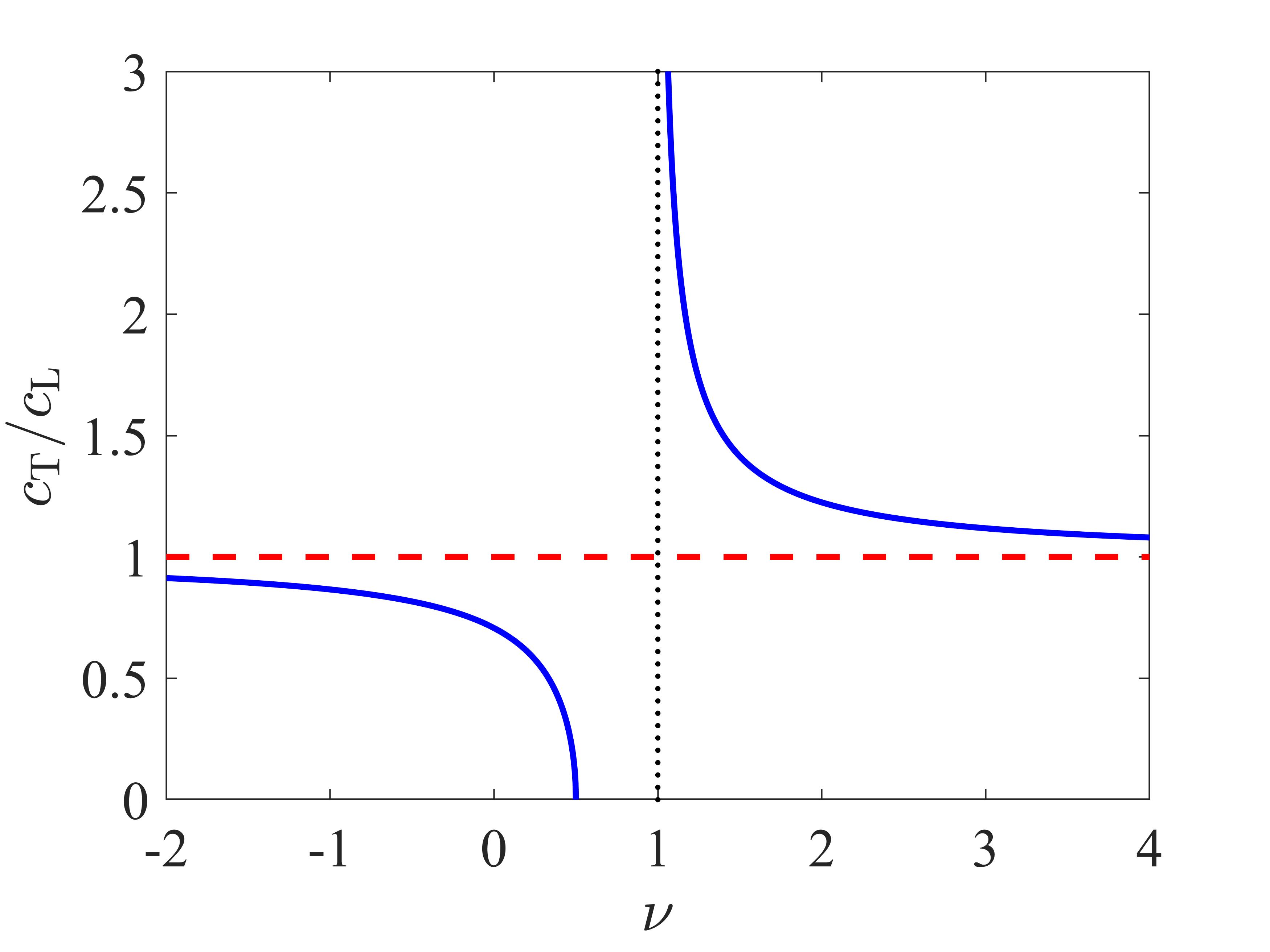}
\caption{\label{fig:F2} The ratio between transverse wave speed and longitudinal wave speed in terms Poisson's ratio under the assumption of isotropy. The red dashed line represents the equal-speed criterion of $c_\text{T}/c_\text{L} = 1$.}
\end{figure}

We consider the frequency-independent wave speeds at the long-wavelength limit, i.e., the quasi-static limit, in an elastic material with isotropy, which is often desirable in applications~\cite{Chen2020Isotropic}. The elasticity can be described by two independent constants~\cite{Landau1970}. For simplicity of discussion, here we use shear modulus $\mu$ and Poisson's ratio $\nu$.\\
The velocities of transverse wave $c_\text{T}$ and longitudinal wave $c_\text{L}$ are:\\
\begin{equation}
    \begin{split}
    c_\text{T} &=\sqrt{\frac{\mu}{\rho}}\\
    c_\text{L} &=\sqrt{\frac{2\mu(1-\nu)}{\rho(1-2\nu)}}\\ 
    \end{split}
\end{equation}
where $\rho$ is the mass density of the material. The wave speed ratio is:\\
\begin{equation}
    c_\text{T}/c_\text{L} = \sqrt{\frac{1-2\nu}{2(1-\nu)}} = \sqrt{1-\frac{1}{2(1-\nu)}}.
\label{ratio}\end{equation} \\
The classical upper bound given by Landau \& Lifshitz~\cite{Landau1970} is
\begin{equation}
c_\text{T}/c_\text{L} < \sqrt{3}/2 < 1.
\end{equation}
This bound is based on the limits of $\nu$:
\begin{equation}
-1 < \nu < 1/2
\end{equation}
which is obtained by requiring both bulk and shear moduli to be positive.\\
With modern metamaterial concepts in mind, we now know that it is possible to achieve  negative bulk modulus (e.g., post-buckling structures, active materials with energy source / sink, etc.). To explore the possibilities in the most general case, we assume that the ratio $c_\text{T}/c_\text{L}$ defined in Eqn. (\ref{ratio}) has no other constraints what so ever. In Fig. \ref{fig:F2},  we plot possible speed ratio values by varying $\nu$. It is clear from the graph that we can actually achieve $c_\text{T}/c_\text{L}>1$ for any $\nu>1$, which indeed implies a negative bulk modulus for any positive shear modulus.\\
However, the speed ratio can only asymptotically approach $1$ at the limits of $\nu \rightarrow \pm \infty$. We note that, while the concept of ``infinite Poisson's ratio'' may be demonstrated as a dynamic-equivalency effective property in locally resonant metamaterials~\cite{ding2007metamaterial}, it is inherently frequency-dependent and narrow-band. Also, the resonance will make both group velocities vanish (i.e. the flat band). In such case, we still cannot have a propagating non-praxial phononic spin. Therefore, for frequency-independent properties in the long-wavelength limit, while normal materials always have $c_\text{T}<c_\text{L}$ and exotic metamaterials may accomplish $c_\text{T}>c_\text{L}$, it is not possible to satisfy the \textit{equal-speed criterion}, $c_\text{T}=c_\text{L}$, with isotropy.\\

\clearpage
\section{Wave speeds in anisotropic materials}

We follow the notations and conventions used in \cite{carcione2007wave} for discussions below. The equation of motion without body force is,
\begin{equation}
    \boldsymbol{\nabla\cdot\sigma}=\rho\Ddot{\boldsymbol{u}},
\label{P1}\end{equation}
where $\boldsymbol{\sigma}$ is stress vector in Voigt notation, $\rho$ is density, $\boldsymbol{u}$ is displacement, $\Ddot{\boldsymbol{u}}$ denotes second order time derivatives of displacement and $\boldsymbol{\nabla}$ is spatial gradient vector in Voigt notation,\\
\begin{equation}
    \bm{\nabla}=\left( \begin{array}{cccccc}
    \partial_1 & 0 & 0 & 0 & \partial_3 & \partial_2\\
    0 & \partial_2 & 0 & \partial_3 & 0 & \partial_1\\
    0 & 0 & \partial_3 & \partial_2 & \partial_1 & 0
\end{array} 
\right ).
\end{equation}
The stress vector can be obtained through constitutive relation in terms of displacement by,
\begin{equation}
    \boldsymbol{\sigma}=\boldsymbol{C \cdot \nabla}^\text{T}\boldsymbol{\cdot u},
\label{P2}\end{equation}
where $\bm{C}$ is rank-2 stiffness tensor in Voigt notation.\\
The general solution of plane wave is
\begin{equation}
    \boldsymbol{u}=\boldsymbol{\tilde{u}} \exp[\text{i} (\boldsymbol{k\cdot r}-\omega t)],
\label{eq:71}\end{equation}
where $\boldsymbol{\tilde{u}}$ is the complex-valued displacement vector, $\boldsymbol{k}$ is the wave vector and $\boldsymbol{r}$ is the position vector. The wave propagates along the $\boldsymbol{\tilde{k}}$-direction,
\begin{equation}
    \boldsymbol{\tilde{k}}=\frac{\bm{k}}{k}=l_1\boldsymbol{e}_1+l_2\boldsymbol{e}_2+l_3\boldsymbol{e}_3,
\label{eq:72}\end{equation}
with $l_1,l_2,l_3$ are direction cosines of the wave vector and $k=|\bm{k}|$ is the wavenumber.\\
With the definition of $\textbf{L}$,
\begin{equation}
    \textbf{L}=\left( \begin{array}{cccccc}
    l_1 & 0 & 0 & 0 & l_3 & l_2\\
    0 & l_2 & 0 & l_3 & 0 & l_1\\
    0 & 0 & l_3 & l_2 & l_1 & 0
\end{array} 
\right ),
\end{equation}
The spatial derivative operator is equivalent to,
\begin{equation}
    \boldsymbol{\nabla} \leftrightarrow \text{i} k \textbf{L}.
\label{eq:73}\end{equation}
Therefore, we can substitute Eqs. \eqref{P2}-\eqref{eq:71} in \eqref{P1} by \eqref{eq:72}-\eqref{eq:73} .
\begin{equation}
    k^2 \boldsymbol{\Gamma \cdot u}=\rho \omega^2 \boldsymbol{u} \quad \text{or} \quad k^2 \Gamma_{ij}u_j=\rho \omega^2 u_i,
\label{eq:74}\end{equation}
where
\begin{equation}
    \boldsymbol{\Gamma}=\textbf{L} \boldsymbol{\cdot} \textbf{C} \boldsymbol{\cdot} \textbf{L}^T \quad \text{or} \quad \Gamma_{ij}=L_{iI}C_{IJ}L_{Jj},
\end{equation}
is the Kelvin-Christoffel matrix with elastic stiffness $C_{IJ}$ in Voigt notation $(I,J=1,2,3...6)$. \comment{Here, the relation between the rank-4 elastic tensor $\mathbb{C}_{ijkl}$ and the rank-2 tensor $C_{IJ}$ is the mappings from $(i,j)$ to $I$, and $(k,l)$ to $J$. For example, the relation between subscripts $ij\rightarrow I$ is $(11,22,33,23,13,12)\rightarrow(1,2,3,4,5,6)$.}\\
More explicitly, the matrix \textbf{C} relates the stress and strain components in the following way: 
\begin{equation}
    \left( \begin{array}{c}
      \sigma_{xx}  \\ \sigma_{yy}  \\ \sigma_{zz} \\ \sigma_{yz} \\ \sigma_{xz} \\ \sigma_{xy}
    \end{array}\right)=\left( \begin{array}{cccccc}
    C_{11} & C_{12} & C_{13} & C_{14} & C_{15} & C_{16}\\
     & C_{22} & C_{23} & C_{24} & C_{25} & C_{26}\\
     &  & C_{33} & C_{34} & C_{35} & C_{36}\\
     &  &  & C_{44} & C_{45} & C_{46}\\
     & \text{\large Sym} &  &  & C_{55} & C_{56}\\
     &  &  &  &  & C_{66}\end{array} \right )\boldsymbol{\cdot}
     \left( \begin{array}{c}
      \epsilon_{xx}  \\ \epsilon_{yy}  \\ \epsilon_{zz} \\ 2\epsilon_{yz} \\ 2\epsilon_{xz} \\ 2\epsilon_{xy}
    \end{array}\right).
\end{equation}
Denoting the phase velocity by $\boldsymbol{c}$:
\begin{equation}
    \boldsymbol{c}=c \tilde{\boldsymbol{k}} \quad \text{with} \quad c=\frac{\omega}{k},
\end{equation}
Eq. \eqref{eq:74} can rewritten as
\begin{equation}
    \boldsymbol{\Gamma \cdot u}-\rho c^2 \boldsymbol{u}=(\boldsymbol{\Gamma}-\rho c^2\textbf{I}) \boldsymbol{\cdot u}=\boldsymbol{0}.
\label{eq:75}\end{equation}
Hence, the 3 eigenvalues of the 3-by-3 Kelvin-Christoffel matrix $\boldsymbol{\Gamma}$ have a one-to-one correspondence with the 3 wave speeds in the anisotropic solid.\\
We note that all the derivations are based on the Cartesian coordinate system, but it can be generalized to other coordinates in 3D, such as the cylindrical coordinates.\\

\clearpage
\subsection{General (Triclinic)}
For a general anisotropic (a.k.a Triclinic) case, we have
\begin{equation}
    \textbf{C}=\left( \begin{array}{cccccc}
    C_{11} & C_{12} & C_{13} & C_{14} & C_{15} & C_{16}\\
     & C_{22} & C_{23} & C_{24} & C_{25} & C_{26}\\
     &  & C_{33} & C_{34} & C_{35} & C_{36}\\
     &  &  & C_{44} & C_{45} & C_{46}\\
     & \text{\large Sym} &  &  & C_{55} & C_{56}\\
     &  &  &  &  & C_{66}
\end{array} 
\right ).
\end{equation}
Then the expressions of each elements in matrix $\boldsymbol{\Gamma}$ are,
\begin{equation}\begin{split}
    &\Gamma_{11}=C_{11}l_1^2+C_{66}l_2^2+C_{55}l_3^2+2C_{56}l_2l_3+2C_{15}l_1l_3+2C_{16}l_1l_2,\\
    &\Gamma_{22}=C_{66}l_1^2+C_{22}l_2^2+C_{44}l_3^2+2C_{24}l_2l_3+2C_{46}l_1l_3+2C_{26}l_1l_2,\\
    &\Gamma_{33}=C_{55}l_1^2+C_{44}l_2^2+C_{33}l_3^2+2C_{34}l_2l_3+2C_{35}l_1l_3+2C_{45}l_1l_2,\\
    &\begin{split}\Gamma_{12}=&C_{16}l_1^2+C_{26}l_2^2+C_{45}l_3^2+(C_{46}+C_{25})l_2l_3+(C_{14}+C_{56})l_1l_3+(C_{12}+C_{66})l_1l_2,\end{split}\\
    &\begin{split}\Gamma_{13}=&C_{15}l_1^2+C_{46}l_2^2+C_{35}l_3^2+(C_{45}+C_{36})l_2l_3+(C_{13}+C_{55})l_1l_3+(C_{14}+C_{56})l_1l_2,\end{split}\\
    &\begin{split}\Gamma_{23}=&C_{56}l_1^2+C_{24}l_2^2+C_{34}l_3^2+(C_{44}+C_{23})l_2l_3+(C_{36}+C_{45})l_1l_3+(C_{25}+C_{46})l_1l_2.\end{split}
\end{split}
\label{P3}\end{equation}
Without loss of generality, we consider a wave propagating along a specific direction, e.g., the $z-$axis:
\begin{equation}
    l_1=0, \ l_2=0, \ l_3=1.
\label{P4}\end{equation}
For other directions with different $(l_1,l_2,l_3)$, the analysis can be obtained similarly.\\
Plug Eq. \eqref{P4} into \eqref{P3}, the Kelvin-Christoffel matrix becomes,
\begin{equation}\begin{split}
    &\Gamma_{11}=C_{55},\\
    &\Gamma_{22}=C_{44},\\
    &\Gamma_{33}=C_{33},\\
    &\Gamma_{12}=C_{45},\\
    &\Gamma_{13}=C_{35},\\
    &\Gamma_{23}=C_{34}.
\end{split}
\end{equation}
Therefore, the governing equation \eqref{eq:75} becomes,
\begin{equation}\begin{split}
     \boldsymbol{0}&=(\boldsymbol{\Gamma}-\rho c^2\textbf{I}) \boldsymbol{\cdot u}\\
     &=\left( \begin{array}{ccc}
        C_{55}-\rho c^2 & C_{45} & C_{35}\\
        C_{45} & C_{44}-\rho c^2 & C_{34}\\
        C_{35} & C_{34} & C_{33}-\rho c^2
    \end{array} \right )
    \boldsymbol{\cdot} \left( \begin{array}{c}
      u_1  \\ u_2 \\  u_3 
    \end{array}\right).\\
\end{split}
\label{eq:79}\end{equation}
Hence, the 3 wave velocities depend on the 3 eigenvalues of matrix $\boldsymbol{\Gamma}$:
\begin{equation}
    c_j=\sqrt{\lambda_j/\rho}
\end{equation}
where $\lambda_j$ for $j=1,2,3$ are eigenvalues of $\boldsymbol{\Gamma}$.\\
Eq. (\ref{eq:79}) may seem to indicate that all three waves are coupled together, preventing any spin polarizations, as we can no longer specify arbitrary phase differences freely.\\
However, we note that $\boldsymbol{\Gamma}$ is a real symmetric matrix, so it is diagonalizable by an orthogonal matrix $\textbf{Q}$: 
\begin{equation}
\textbf{D} = \textbf{Q} \boldsymbol{\cdot} \boldsymbol{\Gamma} \boldsymbol{\cdot} \textbf{Q}^\text{T} =\left( \begin{array}{ccc}
        \lambda_1 & 0 & 0\\
        0 & \lambda_2 & 0\\
        0 & 0 & \lambda_3
    \end{array} \right ).
\label{diagonalization}\end{equation}
Since $\textbf{Q}$ is orthogonal, we have $\det{(\textbf{Q})} = \pm 1$. Then we can define:
\begin{equation}\\
\textbf{R} = \begin{cases}
\textbf{Q}  &\text{if} \quad \det{(\textbf{Q})} = 1\\
-\textbf{Q} &\text{if} \quad \det{(\textbf{Q})} = -1
\end{cases}
\end{equation}
It is apparent that $\textbf{R}$ also diagonalizes $\boldsymbol{\Gamma}$. Since $\textbf{R}$ is an orthogonal matrix with positive unitary determinant, it must be a rotation matrix in the three-dimensional Euclidean space (i.e., a representation of the special orthogonal group SO$(3)$). Therefore, we know that, by rigid rotation of the material (or equivalently, rotating the coordinate system), we can always find the directions in which all three waves with orthogonal displacement fields are decoupled from each other. Thus, as long as $\boldsymbol{\Gamma}$ has degenerate (equal) eigenvalues, we still have the freedom to use phase differences between any two equal-speed modes to create a well-defined propagating spin angular momentum of the traveling wave.\\
In general, none of those decoupled modes needs to be parallel or perpendicular to the $z$-direction. Since the matrix $\boldsymbol{\Gamma}$ is built on the assumption of propagation along the $z$-direction: $\boldsymbol{\tilde{k}} = (l_1=0,l_2=0,l_3=1)$, we may not have pure longitudinal or pure transverse waves as independent and decoupled modes any more.\\
Next, we consider the special case of the ``ultimate" \textit{equal-speed criterion} for all three wave speeds to be the same. This needs all three eigenvalues of $\boldsymbol{\Gamma}$ to be equal ($\lambda=\lambda_1=\lambda_2 = \lambda_3$), then we have:\\
\begin{equation}
\textbf{D} = \lambda\textbf{I}  \quad \Rightarrow \quad \boldsymbol{\Gamma} = \textbf{Q}^\text{T} \textbf{D} \textbf{Q} 
= \lambda\textbf{Q}^\text{T} \textbf{Q} = \lambda\textbf{I},
\end{equation}
since $\textbf{Q}^\text{T}$\textbf{Q} = \textbf{I} always holds for any orthogonal matrix \textbf{Q}. Therefore, $\boldsymbol{\Gamma}$ must be a \textit{scalar matrix} of the form $\lambda\textbf{I}$. Consequently, the requirements become: \\
\begin{equation}\label{ultimate}
C_{33}=C_{44}=C_{55} \quad \text{and} \quad C_{35}=C_{45}=C_{34}=0.
\end{equation}

\clearpage
\subsection{Orthotropic}
The stiffness matrix for the orthotropic case is,
\begin{equation}
    \textbf{C}(\rm Orthotropic)=\left( \begin{array}{cccccc}
    C_{11} & C_{12} & C_{13} & 0 & 0 & 0\\
     & C_{22} & C_{23} & 0 & 0 & 0\\
     &  & C_{33} & 0 & 0 & 0\\
     &  &  & C_{44} & 0 & 0\\
     & \text{\large Sym} &  &  & C_{55} & 0\\
     &  &  &  &  & C_{66}
\end{array} 
\right ), \quad (9 \ \rm constants).
\end{equation}
For the wave propagating along $z-$direction,
\begin{equation}
    l_1=0, \ l_2=0, \ l_3=1.
\end{equation}
Then, the corresponding Kelvin-Christoffel matrix becomes,
\begin{equation}\begin{split}
    &\Gamma_{11}=C_{55},\\
    &\Gamma_{22}=C_{44},\\
    &\Gamma_{33}=C_{33},\\
    &\Gamma_{12}=0,\\
    &\Gamma_{13}=0,\\
    &\Gamma_{23}=0.
\end{split}
\end{equation}
Therefore, the governing equation \eqref{eq:75} becomes,
\begin{equation}\begin{split}
     \boldsymbol{0}&=(\boldsymbol{\Gamma}-\rho c^2\textbf{I}) \boldsymbol{\cdot u}\\
     &=\left( \begin{array}{ccc}
        C_{55}-\rho c^2 & 0 & 0\\
         0 & C_{44}-\rho c^2 & 0\\
        0 & 0 & C_{33}-\rho c^2
    \end{array} \right )
    \boldsymbol{\cdot} \left( \begin{array}{c}
      u_1  \\ u_2 \\  u_3 
    \end{array}\right).
\end{split}
\end{equation}
It is noticed that all waves are decoupled. Considering that the wave propagates along $z-$direction, the requirement of equal-speed propagation is,
\begin{equation}
    C_{33}=C_{55} \quad \text{in the $xz$-plane}
\end{equation}
\begin{equation}
    C_{33}=C_{44} \quad \text{in the $yz$-plane}
\end{equation}

\comment{
For the waves propagating along $x$- and $y$-directions, we can set the direction cosines accordingly, \\
\begin{equation}
    l_1=1, \ l_2=0, \ l_3=0 \quad \text{for the $x$-direction}
\end{equation}
 and
\begin{equation}
    l_1=0, \ l_2=1, \ l_3=0 \quad \text{for the $y$-direction}
\end{equation}

Similarly, we can obtain the $\boldsymbol{\Gamma}$ matrices and then the corresponding governing equations. Because of the property of orthotropic material, we will have very similar requirements of the rolling wave.\\

For waves propagating along the $x$-direction, the requirements are
\begin{equation}
    C_{11}=C_{66}
\end{equation}
in the $xy$-plane and
\begin{equation}
    C_{11}=C_{55}
\end{equation}
in the $xz$-plane.\\

For waves propagating along the $x$-direction, the requirements are
\begin{equation}
    C_{22}=C_{66}
\end{equation}
in the $xy$-plane and
\begin{equation}
    C_{22}=C_{44}
\end{equation}
in the $yz$-plane.
}

\clearpage
\subsection{Transversely Isotropic}
The stiffness matrix for transversely isotropic material is defined by 5 independant elastic constants,
\begin{equation}
    \textbf{C}=\left( \begin{array}{cccccc}
    C_{11} & C_{12} & C_{13} & 0 & 0 & 0\\
     & C_{11} & C_{13} & 0 & 0 & 0\\
     &  & C_{33} & 0 & 0 & 0\\
     &  &  & C_{44} & 0 & 0\\
     & \text{\large Sym} &  &  & C_{44} & 0\\
     &  &  &  &  & C_{66}
\end{array} 
\right )
\end{equation}
where $C_{66}$ is not independent, 
\begin{equation}
C_{66}=\frac{C_{11}-C_{12}}{2}.
\end{equation}
The transversely isotropic material defined above has isotropic property in the $xy$-plane and anisotropic property along $z$-direction. \\
For the wave propagating along $z$-direction,
\begin{equation}
    l_1=0, \ l_2=0, \ l_3=1.
\end{equation}
Then, the corresponding Kelvin-Christoffel matrix becomes,
\begin{equation}\begin{split}
    &\Gamma_{11}=C_{44},\\
    &\Gamma_{22}=C_{44},\\
    &\Gamma_{33}=C_{33},\\
    &\Gamma_{12}=0,\\
    &\Gamma_{13}=0,\\
    &\Gamma_{23}=0.
\end{split}
\end{equation}
Therefore, the governing equation \eqref{eq:75} becomes,
\begin{equation}\begin{split}
     \boldsymbol{0}&=(\boldsymbol{\Gamma}-\rho c^2\textbf{I}) \boldsymbol{\cdot u}\\
     &=\left( \begin{array}{ccc}
        C_{44}-\rho c^2 & 0 & 0\\
        0 & C_{44}-\rho c^2 & 0\\
        0 & 0 & C_{33}-\rho c^2
    \end{array} \right )
    \boldsymbol{\cdot} \left( \begin{array}{c}
      u_1  \\ u_2 \\  u_3 
    \end{array}\right).
\end{split}
\label{eq:76}\end{equation}
All waves are decoupled. The longitudinal wave relates to $C_{33}$ and both shear waves relate to $C_{44}$. Therefore, the requirement of equal-speed propagation along the $z$-direction is,
\begin{equation}
    C_{33}=C_{44}.
\end{equation}

\clearpage
\subsection{Cubic}
The stiffness matrix for cubic material is,
\begin{equation}
    \textbf{C}(\rm Cubic)=\left( \begin{array}{cccccc}
    C_{11} & C_{12} & C_{12} & 0 & 0 & 0\\
     & C_{11} & C_{12} & 0 & 0 & 0\\
     &  & C_{11} & 0 & 0 & 0\\
     &  &  & C_{44} & 0 & 0\\
     & \text{\large Sym} &  &  & C_{44} & 0\\
     &  &  &  &  & C_{44}
\end{array} 
\right ), \quad \left(3 \ \rm constants\right ).
\end{equation}
All 3 directions along the coordinate axes are same due to the cubic symmetry. Therefore, it is sufficient to analyze the $x$-direction only.\\
For the wave propagating along the $x$-direction, the direction cosines are
\begin{equation}
    l_1=1, \ l_2=0, \ l_3=0.
\end{equation}
Then, the corresponding Kelvin-Christoffel matrix becomes,
\begin{equation}\begin{split}
    \Gamma_{11}&=C_{11},\\
    \Gamma_{22}&=C_{44},\\
    \Gamma_{33}&=C_{44},\\
    \Gamma_{12}=\Gamma_{13}&=\Gamma_{23}=0.\\
\end{split}
\end{equation}
Therefore, the governing equation \eqref{eq:75} becomes,
\begin{equation}\begin{split}
     \boldsymbol{0}&=(\boldsymbol{\Gamma}-\rho c^2\textbf{I}) \boldsymbol{\cdot u}\\
     &=\left( \begin{array}{ccc}
        C_{11}-\rho c^2 & 0 & 0\\
         0 & C_{44}-\rho c^2 & 0\\
        0 & 0 & C_{44}-\rho c^2
    \end{array} \right )
    \boldsymbol{\cdot} \left( \begin{array}{c}
      u_1  \\ u_2 \\  u_3 
    \end{array}\right).
\end{split}
\label{eq:86}\end{equation}
Here, $u_1$ represents longitudinal wave and $u_2,u_3$ represent transverse waves. Therefore, from Eq. \eqref{eq:86}, we obtain
\begin{equation}
        c_\text{L}=\sqrt{\frac{C_{11}}{\rho}},\ \ c_\text{T}=\sqrt{\frac{C_{44}}{\rho}}.
\end{equation}
Thus, to generate the rolling  wave in $x$-direction, we only need to have,
\begin{equation}
    C_{11}=C_{44}.
    \label{cubic_criterion}
\end{equation}

\clearpage
\subsection{Plane Strain}
The plane strain condition of $xz$-plane for general anistropic material is,
\begin{equation}
    \boldsymbol{\epsilon}=\left( \begin{array}{c}
      \epsilon_{xx}  \\ \epsilon_{yy}  \\ \epsilon_{zz} \\ 2\epsilon_{yz} \\ 2\epsilon_{xz} \\ 2\epsilon_{xy}
    \end{array}\right)
    =\left( \begin{array}{c}
      \epsilon_{xx}  \\ 0  \\ \epsilon_{zz} \\ 0 \\ 2\epsilon_{xz} \\ 0
    \end{array}\right).
\end{equation}
The general elastic stiffness tensor in Voigt notation is,
\begin{equation}
    \textbf{C}=\left( \begin{array}{cccccc}
    C_{11} & C_{12} & C_{13} & C_{14} & C_{15} & C_{16}\\
     & C_{22} & C_{23} & C_{24} & C_{25} & C_{26}\\
     &  & C_{33} & C_{34} & C_{35} & C_{36}\\
     &  &  & C_{44} & C_{45} & C_{46}\\
     & \text{\large Sym} &  &  & C_{55} & C_{56}\\
     &  &  &  &  & C_{66}
\end{array} 
\right ).
\end{equation}
Then, by substituting into the constitutive relation $\boldsymbol{\sigma}=\boldsymbol{C\cdot\epsilon}$, we have
\begin{equation}
    \left( \begin{array}{c}
      \sigma_{xx}  \\ \sigma_{yy}  \\ \sigma_{zz} \\ \sigma_{yz} \\ \sigma_{xz} \\ \sigma_{xy}
    \end{array}\right)=\left( \begin{array}{cccccc}
    C_{11} & C_{12} & C_{13} & C_{14} & C_{15} & C_{16}\\
     & C_{22} & C_{23} & C_{24} & C_{25} & C_{26}\\
     &  & C_{33} & C_{34} & C_{35} & C_{36}\\
     &  &  & C_{44} & C_{45} & C_{46}\\
     & \text{\large Sym} &  &  & C_{55} & C_{56}\\
     &  &  &  &  & C_{66}\end{array} \right )\boldsymbol{\cdot}
     \left( \begin{array}{c}
      \epsilon_{xx}  \\ 0  \\ \epsilon_{zz} \\ 0 \\ 2\epsilon_{xz} \\ 0
    \end{array}\right).
\end{equation}
Therefore, for the 2D plane strain cases, we only need to consider the following stress components,
\begin{equation}
    \left( \begin{array}{c}
      \sigma_{xx}  \\ \sigma_{zz} \\ \sigma_{xz}
    \end{array}\right)=\left( \begin{array}{ccc}
    C_{11} & C_{13} & C_{15}\\
    C_{31} & C_{33} & C_{35}\\
    C_{51} & C_{53} & C_{55}\end{array} \right )\boldsymbol{\cdot}
     \left( \begin{array}{c}
      \epsilon_{xx}  \\ \epsilon_{zz} \\ 2\epsilon_{xz}
    \end{array}\right).
\end{equation}
The governing equation of plane strain situation becomes,
\begin{equation}
     (\boldsymbol{\Gamma}-\rho c^2\textbf{I}) \boldsymbol{\cdot u}=\boldsymbol{0},
\end{equation}
where $\boldsymbol{\Gamma}$ is the Kelvin-Christoffel matrix in $xz$-plane,
\begin{equation}
    \boldsymbol{\Gamma}=\left( \begin{array}{cc}
        \Gamma_{11} & \Gamma_{13}\\
        \Gamma_{13} & \Gamma_{33}
    \end{array} \right ),
\end{equation}
which is from Eq. \eqref{P3}.\\
For $z$-direction propagating wave,
\begin{equation}
    l_1=0,\ \ l_2=0,\ \ l_3=1.
\end{equation}
The corresponding Kelvin-Christoffel matrix becomes,
\begin{equation}\begin{split}
    &\Gamma_{11}=C_{55},\\
    &\Gamma_{33}=C_{33},\\
    &\Gamma_{13}=C_{35}.
\end{split}
\end{equation}
The governing equation of plane strain situation becomes,
\begin{equation}\begin{split}
     \boldsymbol{0}&=(\boldsymbol{\Gamma}-\rho c^2\textbf{I}) \cdot\boldsymbol{u}\\
     &=\left( \begin{array}{cc}
        C_{55}-\rho c^2 & C_{35}\\
        C_{35} & C_{33}-\rho c^2
    \end{array} \right )
    \boldsymbol{\cdot} \left( \begin{array}{c}
      u_1 \\  u_3 
    \end{array}\right).
\end{split}
\end{equation}
Hence, for $u_1$ and $u_3$ to be decoupled and to propagate at the same wave speed, the requirements are $C_{35} = 0$ and $C_{55} = C_{33}$.

\clearpage
\section{The Square Lattice}
\begin{figure}[htb]
\centering
\includegraphics[scale=0.5]{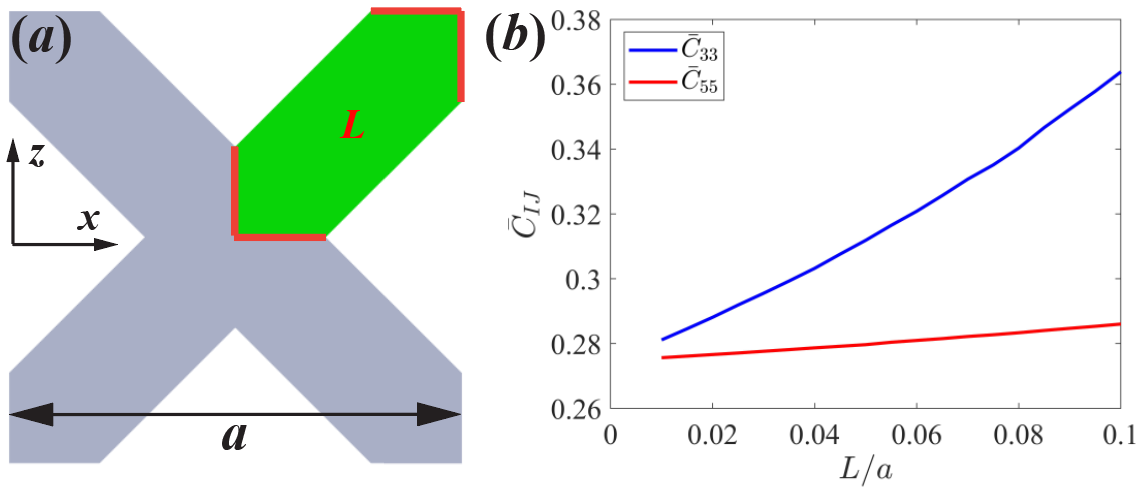}
\caption{\label{fig:sq} (a) The unit cell of a 2D square lattice. It is constructed by taking mirror images of the green quarter. All red straight line segments are of length $L$. The structure illustrated here is with $L/a=0.2$. The assumed propagation direction along the $z$-axis corresponds to the diagonal (i.e. 45 degree) direction in the square lattice (b) The non-dimensional elastic constants results with varying $L/a$.}
\end{figure}\\
\noindent Some previous studies~\cite{phani2006,wang2015locally} used the Bloch-wave formulation to calculate dispersion relations for the square lattice of beams. The band structures presented in those studies seem to indicate nearly equal-speed propagation at the low-frequency long-wavelength limit in the diagonal (45 degree) direction of the square.\\
Unfortunately, our calculations, as presented in Fig.\,\ref{fig:sq}, show that the square lattice can only asymptotically approach the limit of $c_\text{L} = c_\text{T}$ (equivalent to $C_{33}=C_{55}$ according to Eq. (11a) in the main text) when the beam width $\rightarrow 0$. This fact renders square lattices with finite bending stiffness impractical for hosting rolling waves.\\
For completeness, we also perform the band structure calculations to match the results in \cite{phani2006,wang2015locally} as well as the zoom-in calculations in the long wavelength limit. We note that the ``slenderness ratio" defined in both studies is equivalent to $(a/\sqrt{2})\sqrt{(E (\sqrt{2}L)^2)/(E\frac{(\sqrt{2}L)^4}{12})} = \sqrt{3}(a/L)$ for the parameters defined in Fig.\,\ref{fig:sq}(a), since we actually have the beam thickness = $\sqrt{2}L$ and the conventional square unit cell size = $a/\sqrt{2}$.

\comment{
The normalized frequency is defined by,
\begin{equation}
    \Omega=\frac{\omega}{\omega_1},\ \ \text{with}\ \ \omega_1=\pi^2\sqrt{\frac{EI}{\rho (\sqrt{2}L)(a/\sqrt{2})^4}}.
\label{norm_freq}
\end{equation}
}

\begin{figure}[htb]
\centering
\includegraphics[scale=0.42]{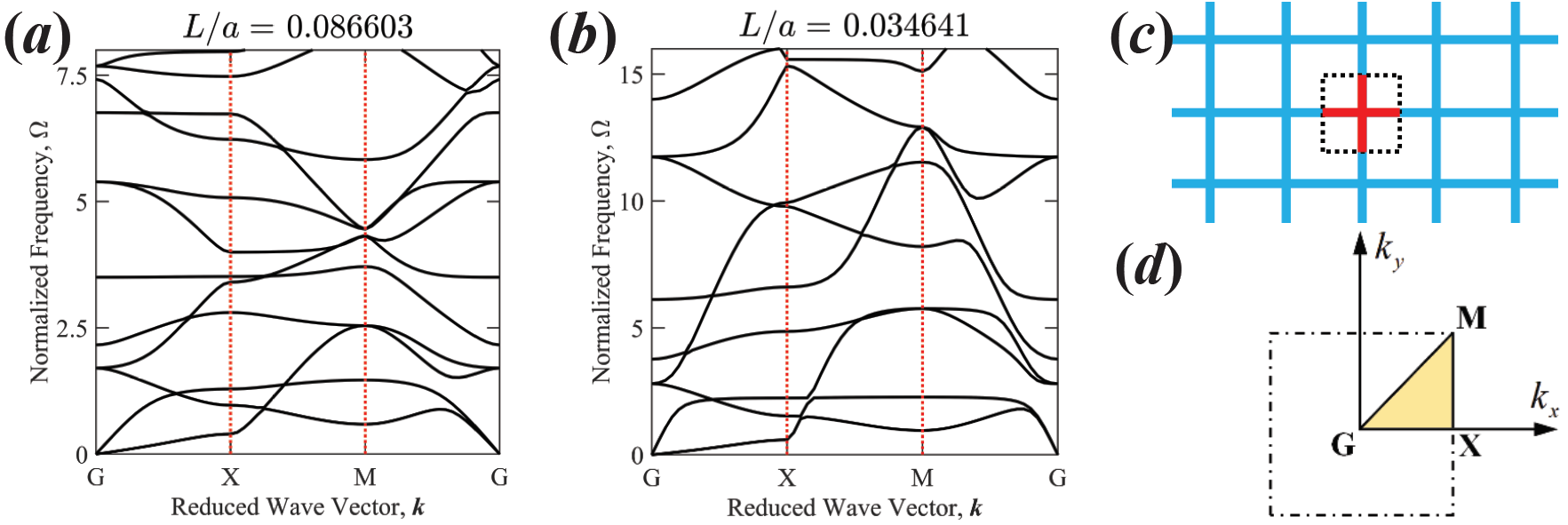}
\caption{Dispersion relations of the square lattice with different beam thickness: (a) Same slenderness ratio of 20 as \cite{wang2015locally}. (b) Same slenderness ratio of 50 as \cite{phani2006}. (c) Schematic of square lattice. (d) The Brillioun zone of square lattice.  Following the same convention in \cite{phani2006}, the normalized frequency is defined as $\Omega=\frac{\omega}{\omega_1}$
with
$\omega_1=\pi^2\sqrt{\frac{EI}{\rho_{A} (\sqrt{2}L)(a/\sqrt{2})^4}}$, where $\rho_{A}$ denotes mass per unit area, or equivalently the ``2D mass density".}
\label{fig:sq1} 
\end{figure}
\begin{figure}[htb]
\centering
\includegraphics[scale=0.42]{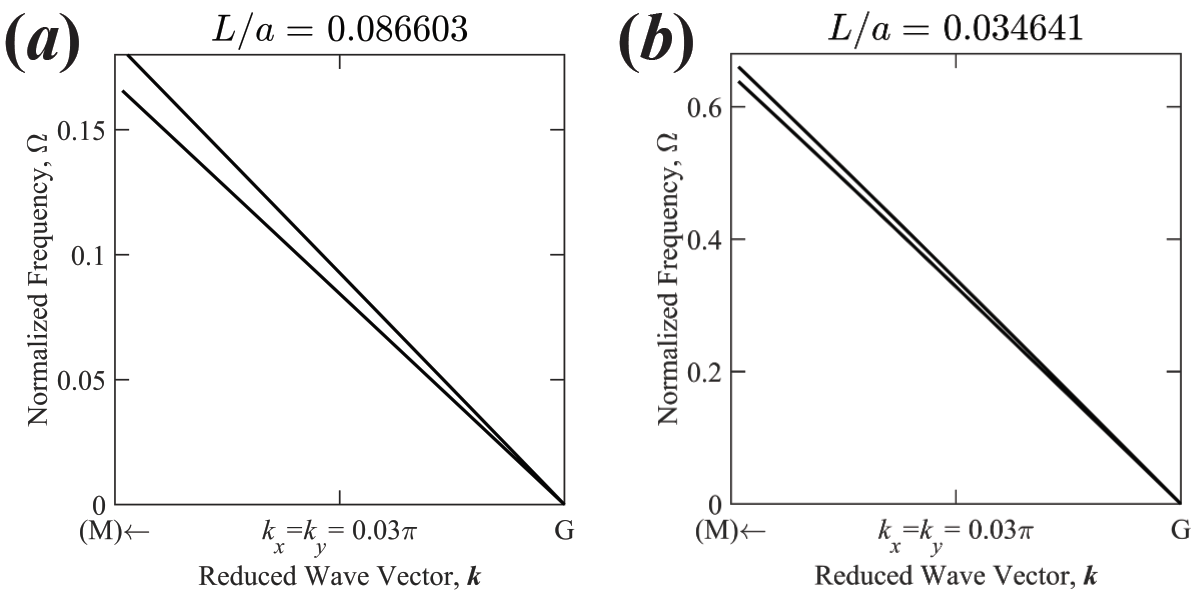}
\caption{Zoomed in Dispersion relations at direction $\text{M}\rightarrow\text{G}$ close to G: (a) Same slenderness ratio of 20 as \cite{wang2015locally}. (b) Same slenderness ratio of 50 as \cite{phani2006}.}
\label{fig:sq2} 
\end{figure}

\clearpage
\section{An alternative 3D design}

\begin{figure}[htb]
\centering
\includegraphics[scale=0.5]{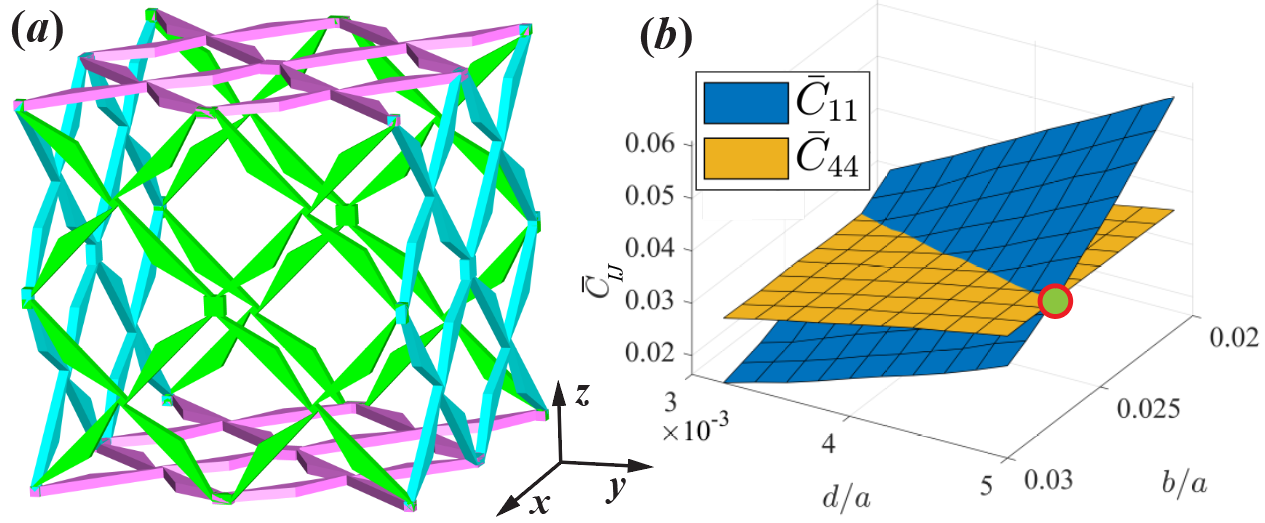}
\caption{\label{fig:3D-alternative} (a) Auxetic unit cell design similar to the geometry used in \cite{buckmann2014three}. Colors are for visual distinction. All six faces are identical. (b) The non-dimensional elastic constants results with varying parameters $b_1/a$ and $b_2/a$. Geometry in (a) corresponds to the circled point in (b).}
\end{figure}

We may also use the structure shown as Fig. 2(c) in the main text to build 3D designs. For example, we can directly use the 2D pattern as each of the six faces of a cube (Fig. \ref{fig:3D-alternative}(a)). This is very similar to the geometry used in \cite{buckmann2014three}, but here we have removed structures inside the unit cube for simplicity. As duly noted in \cite{buckmann2014three}, this 3D geometry itself does not have the symmetry of cubic crystallographic point groups, which are characterised by the four threefold rotation axes along the body diagonals of a cube. Hence, there is no \textit{a priori} guarantee that it will result in an elasticity tensor for cubic symmetry.\\ 
On the other hand, our numerical calculations can confirm that it can still meet the \textit{equal-speed criterion}. Here, each face of the unit cube has a uniform out-of-plane thickness $h_1$ (due to spatial periodicity the metamaterial has a wall thickness of $2h_1$). With the cube edge length being $a$, we fix the parameters as $2h_1=b_3=b_4=0.05a$ and $b_5=0.3221a$. Then we vary $b_1/a$ and $b_2/a$ to calculate the elastic constants in each case. The results are shown in (Fig. \ref{fig:3D-alternative}(b)). The equal-speed  criterion is met at the intersection of two (blue and yellow) surfaces in the parameter space.

\clearpage
\section{Reflection of Rolling Wave at Elastic Boundary}
\subsection{Normal incidence and reflection}
Omitting the time harmonic term $e^{-i\omega t}$ and assuming the principle wave displacement directions coincide with the coordinate system, we consider a general plane wave propagating along $z$-direction incident on a flat surface (the $xy$-plane at $z=0$),
\begin{equation}
    \boldsymbol{u}^\text{I}=\left( \begin{array}{c}
      m^\text{I}  \\ n^\text{I} \\ l^\text{I}
    \end{array}\right)e^{ikz}
\end{equation}
where I denotes the incident wave.\\
The backward reflection wave can be written as,
\begin{equation}
    \boldsymbol{u}^\text{R}=\left( \begin{array}{c}
     m^\text{R}  \\ n^\text{R} \\ l^\text{R}
    \end{array}\right)e^{-ikz}
\end{equation}
where R denotes the reflected wave.\\
The strains can be calculated from displacements by
\begin{equation}
    \epsilon_{ij}=\frac{1}{2}(u_{i,j}+u_{j,i})
\end{equation}
where the comma $``,"$ denotes the derivative operation.\\
The reflection occurs at $z=0$. So we have $e^{ikz} = e^{-ikz} = 1$, and the strain vector in Voigt notation becomes,
\begin{equation}
    \boldsymbol{\epsilon}^\text{I}=\left( \begin{array}{c}
      \epsilon^\text{I}_{xx}  \\ \epsilon^\text{I}_{yy} \\ \epsilon^\text{I}_{zz} \\ 2\epsilon^\text{I}_{yz} \\ 2\epsilon^\text{I}_{xz} \\ 2\epsilon^\text{I}_{xy}
    \end{array}\right)=\left( \begin{array}{c}
       0  \\ 0 \\ l^\text{I} ik \\ n^\text{I} ik  \\ m^\text{I} ik \\ 0
    \end{array}\right),\ \ 
    \boldsymbol{\epsilon}^\text{R}=\left( \begin{array}{c}
      \epsilon^\text{R}_{xx}  \\ \epsilon^\text{R}_{yy} \\ \epsilon^\text{R}_{zz} \\ 2\epsilon^\text{R}_{yz} \\ 2\epsilon^\text{R}_{xz} \\ 2\epsilon^\text{R}_{xy}
    \end{array}\right)=\left( \begin{array}{c}
       0  \\ 0 \\ -l^\text{R} ik \\ -n^\text{R} ik  \\ -m^\text{R} ik \\ 0
    \end{array}\right).
    \label{strain_IR}
\end{equation}
Here we assume the effective orthotropic constitutive relation in Voigt notation:
\begin{equation}
    \left( \begin{array}{c}
      \sigma_{xx}  \\ \sigma_{yy} \\ \sigma_{zz} \\ \sigma_{yz} \\ \sigma_{xz} \\ \sigma_{xy}
    \end{array}\right)=\left( \begin{array}{cccccc}
      C_{11}  &  C_{12} &  C_{13} &  0 &  0 &  0 \\
      C_{12}  &  C_{22} &  C_{23} &  0 &  0 &  0 \\
      C_{13}  &  C_{23} &  C_{33} &  0 &  0 &  0 \\
      0  &  0 &  0 &  C_{44} &  0 &  0 \\
      0  &  0 &  0 &  0 &  C_{55} &  0 \\
      0  &  0 &  0 &  0 &  0 &  C_{66}
    \end{array}\right)\cdot
    \left( \begin{array}{c}
      \epsilon_{xx}  \\ \epsilon_{yy} \\ \epsilon_{zz} \\ 2\epsilon_{yz} \\ 2\epsilon_{xz} \\ 2\epsilon_{xy}
    \end{array}\right).
\end{equation}
\comment{with 
\begin{equation}
    \boldsymbol{\sigma}=\left( \begin{array}{ccc}
      \sigma_{xx}  & \sigma_{xy}  & \sigma_{xz}\\
      \sigma_{yx}  & \sigma_{yy}  & \sigma_{yz}\\
      \sigma_{zx}  & \sigma_{zy}  & \sigma_{zz}  
    \end{array}\right)
\end{equation}
denotes the stress tensor.}\\
It is easy to compute the corresponding stress components for incident and reflect waves.
\begin{equation}
    \left( \begin{array}{c}
      \sigma^\text{I}_{xx}  \\ \sigma^\text{I}_{yy} \\ \sigma^\text{I}_{zz} \\ \sigma^\text{I}_{yz} \\ \sigma^\text{I}_{xz} \\ \sigma^\text{I}_{xy}
    \end{array}\right)=\left( \begin{array}{c}
      C_{13}l^\text{I} ik  \\ C_{23}l^\text{I} ik  \\ C_{33}l^\text{I} ik \\ C_{44}n^\text{I} ik \\ C_{55}m^\text{I} ik \\ 0
    \end{array}\right),\ \ 
    \left( \begin{array}{c}
      \sigma^\text{R}_{xx}  \\ \sigma^\text{R}_{yy} \\ \sigma^\text{R}_{zz} \\ \sigma^\text{R}_{yz} \\ \sigma^\text{R}_{xz} \\ \sigma^\text{R}_{xy}
    \end{array}\right)=\left( \begin{array}{c}
      -C_{13}l^\text{R} ik  \\ -C_{23}l^\text{R} ik  \\ -C_{33}l^\text{R} ik \\ -C_{44}n^\text{R} ik \\ -C_{55}m^\text{R} ik \\ 0
    \end{array}\right).
\end{equation}
For the elastically supported cubic half-space, the boundary conditions are
\begin{equation}
    \sigma^0_{zx}=K_xu^0_x,\ \ \sigma^0_{zy}=K_yu^0_y,\ \ \sigma^0_{zz}=K_zu^0_z.
\end{equation}
where $(K_x,K_y,K_z)$ are components of \underline{distributed stiffness per unit area}~\cite{zhang2017reflection} representing a general elastic foundation supporting the solid surface. The stress and displacement summations are
\begin{equation}
    \sigma^0_{zx}=\sigma^\text{I}_{zx}+\sigma^\text{R}_{zx},\ \ \sigma^0_{zy}=\sigma^\text{I}_{zy}+\sigma^\text{R}_{zy},\ \ \sigma^0_{zz}=\sigma^\text{I}_{zz}+\sigma^\text{R}_{zz}.
\end{equation}
\begin{equation}
    u^0_x=u^\text{I}_x+u^\text{R}_x,\ \ u^0_y=u^\text{I}_y+u^\text{R}_y,\ \ u^0_z=u^\text{I}_z+u^\text{R}_z.
\end{equation}
Substituting the displacement and stress components into elastic boundary condition,
\begin{equation}
    C_{55}m^\text{I} ik-C_{55}m^\text{R} ik=K_x(m^\text{I}+m^\text{R}),
\end{equation}
\begin{equation}
    C_{44}n^\text{I} ik-C_{44}n^\text{R} ik=K_y(n^\text{I}+n^\text{R}),
\end{equation}
\begin{equation}
    C_{33}l^\text{I} ik-C_{33}l^\text{R} ik=K_z(l^\text{I}+l^\text{R}).
\end{equation}
Solving the equation set gives us,
\begin{equation}
    m^\text{R}=\frac{C_{55} ik-K_x}{C_{55} ik+K_x}m^\text{I},
\end{equation}
\begin{equation}
    n^\text{R}=\frac{C_{44} ik-K_y}{C_{44} ik+K_y}n^\text{I},
\end{equation}
\begin{equation}
    l^\text{R}=\frac{C_{33} ik-K_z}{C_{33} ik+K_z}l^\text{I}.
\end{equation}
Because of the requirement of rolling wave ($C_{33}=C_{44}=C_{55}$), the amplitude of reflection wave becomes,
\begin{equation}
    m^\text{R}=R_x m^\text{I},\ \ n^\text{R}=R_y n^\text{I},\ \ l^\text{R}=R_z l^\text{I},
\end{equation}
with 
\begin{equation}
\label{Rxyz}
    R_x=\frac{C_{33} ik-K_x}{C_{33} ik+K_x},\ \ 
    R_y=\frac{C_{33} ik-K_y}{C_{33} ik+K_y},\ \ 
    R_z=\frac{C_{33} ik-K_z}{C_{33} ik+K_z}. 
\end{equation}
Clearly, the reflected wave amplitude depend on the spring stiffness. Moreover, the elastic boundary condition will degenerate into traction free boundary condition when the stiffness $K_j=0$. Then the amplitudes of reflected wave become
\begin{equation}
    m^\text{R}=m^\text{I},\ \ n^\text{R}=n^\text{I},\ \ l^\text{R}=l^\text{I}.
\end{equation}
Similarly, the elastic boundary condition will degenerate into fixed boundary condition when the stiffness $K_j=\infty$. Then, the amplitudes of reflected wave become
\begin{equation}
    m^\text{R}=-m^\text{I},\ \ n^\text{R}=-n^\text{I},\ \ l^\text{R}=-l^\text{I}.
\end{equation}
In both cases above, by the definition of spin given in Eq. (2) of the main text, we can conclude $\boldsymbol{s}^\text{R}=\boldsymbol{s}^\text{I}$, so the spin is unaffected by reflection.\\
Next, we consider in-$xz$-plane waves with $n^{\text{I}}=n^{\text{R}}=0$. For the free-rigid hybrid boundary ($K_x=0$,$K_z=\infty$), we have
\begin{equation}
    m^\text{R}=m^\text{I},\ \ l^\text{R}=-l^\text{I} \quad \Rightarrow \quad \boldsymbol{s}^\text{R}=-\boldsymbol{s}^\text{I}.
\end{equation}
Similarly, for the rigid-free hybrid boundary ($K_x=\infty$, $K_z=0$), we have 
\begin{equation}
    m^\text{R}=-m^\text{I},\ \ l^\text{R}=l^\text{I} \quad \Rightarrow \quad \boldsymbol{s}^\text{R}=-\boldsymbol{s}^\text{I}.
\end{equation}
Thus, both hybrid boundaries will flip the spin for any incident rolling wave. \\

\subsection{Complex-valued amplitude ratio $R_j$}
\begin{figure}[h!]
\centering
\includegraphics[scale=0.6]{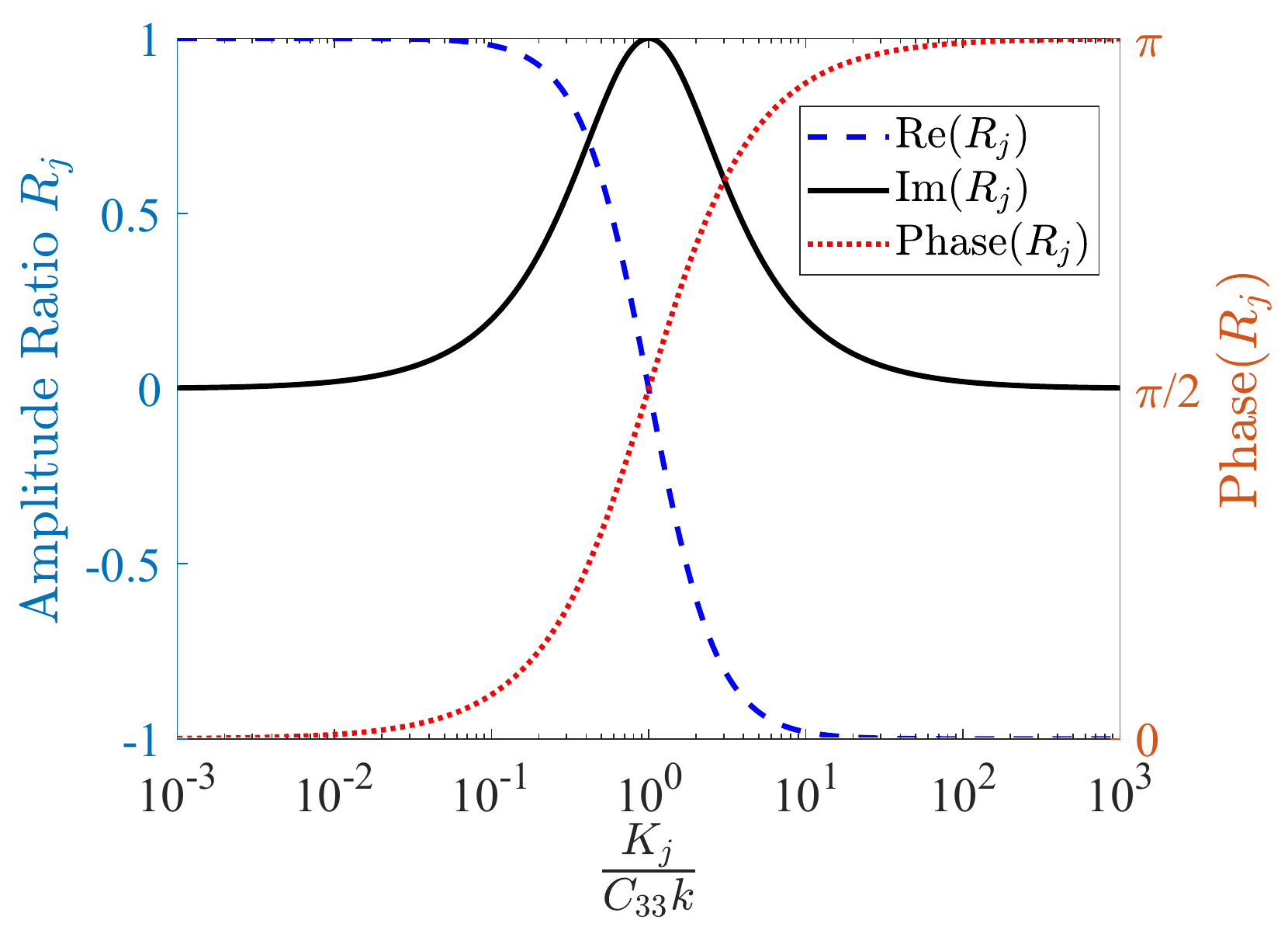}
\caption{\label{fig:Rj} The amplitude ratio $R_j$ between reflect wave and incident wave. }
\end{figure}

The effects of normal reflection can be described by the generalized amplitude ratio in (\ref{Rxyz}):
\begin{equation}
    R_j=\frac{C_{33} ik-K_j}{C_{33} ik+K_j},\ \ j=x,y,z.
\end{equation}
This complex non-dimensional parameter $R_j$ plays a key role between the incident and reflect waves and deserves further analysis. We note that $|R_j|=1$ is consistent with the fact that all wave energy is reflected, and the phase angle $\phi$ represents the phase change during the reflection. Hence, we have \\
\begin{equation}
    R_j = e^{i\phi} \quad \text{with} \quad \tan{\phi}=\frac{2{K_j}/{C_{33}k}}{1-({K_j}/{C_{33}k})^2}
\end{equation}
and it dependents on the boundary-bulk stiffness ratio ${K_j}/{C_{33}k}$. 
By varying this ratio, we plot the real and imaginary parts of $R_j$ as well as the phase angle $\phi$ in Figure \ref{fig:Rj}. These values, with respect to the logarithmic magnitude of the boundary-bulk stiffness ratio, show typical symmetric and anti-symmetric properties.
Therefore, by adjusting the elastic stiffness at the boundary, one can manipulate the spin of the reflection waves.
\\
Although all properties of both the bulk and the reflection surface are assumed to be independent of the incident wave frequency, here the reflection phase change can be frequency-dependent, as the angular wave number $k$ appears in the ratio. The emergence of frequency dependency can be intuitively explained by the role of wavelength during reflection:\\
We note from Eq. (\ref{strain_IR}) that $k$ first appears in the strain calculations since, for a fixed wave displacement amplitude, the strains in the propagation direction, $\epsilon_{zj}$, are actually inversely proportional to the wavelength: Longer wavelength gives rise to a smaller strain and vice versa. Consequently, the stresses, $\sigma_{zj}$, and the force acting on the boundary springs are wavelength-dependent as well. If the boundary is supported by an elastic foundation with finite stiffness per area, $K_j$, we have the following:\\
At the low-frequency and long-wavelength limit, the force per area acting on the boundary, $|\sigma_{zj}| \propto C_{33}k \rightarrow 0$. So, the boundary hardly move, and the incident wave effectively ``sees" a rigid surface;\\
At the high-frequency and short-wavelength limit, the force per area acting on the boundary, $|\sigma_{zj}| \propto C_{33}k \rightarrow \infty$. So, the boundary moves a lot, and the incident wave effectively ``sees" a free surface.\\

\subsection{Time domain simulations}
\begin{figure}[htb]
\centering
\includegraphics[scale=0.8]{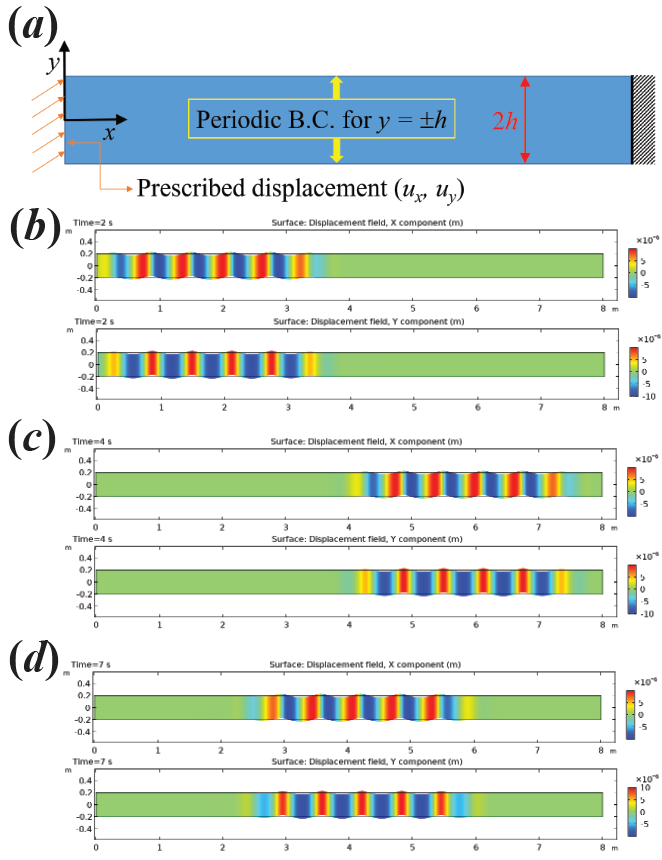}
\caption{\label{fig:SIF3} (a) The schematic of the setup of \textsc{comsol} anisotropic plane model. (b) The displacement fields $u_x$ and $u_y$ at time $t=2$s. (c) The displacement fields $u_x$ and $u_y$ at time $t=4$s. (d) The displacement fields $u_x$ and $u_y$ at time $t=7$s.  }
\end{figure}

\begin{figure}[htb]
\centering
\includegraphics[scale=0.8]{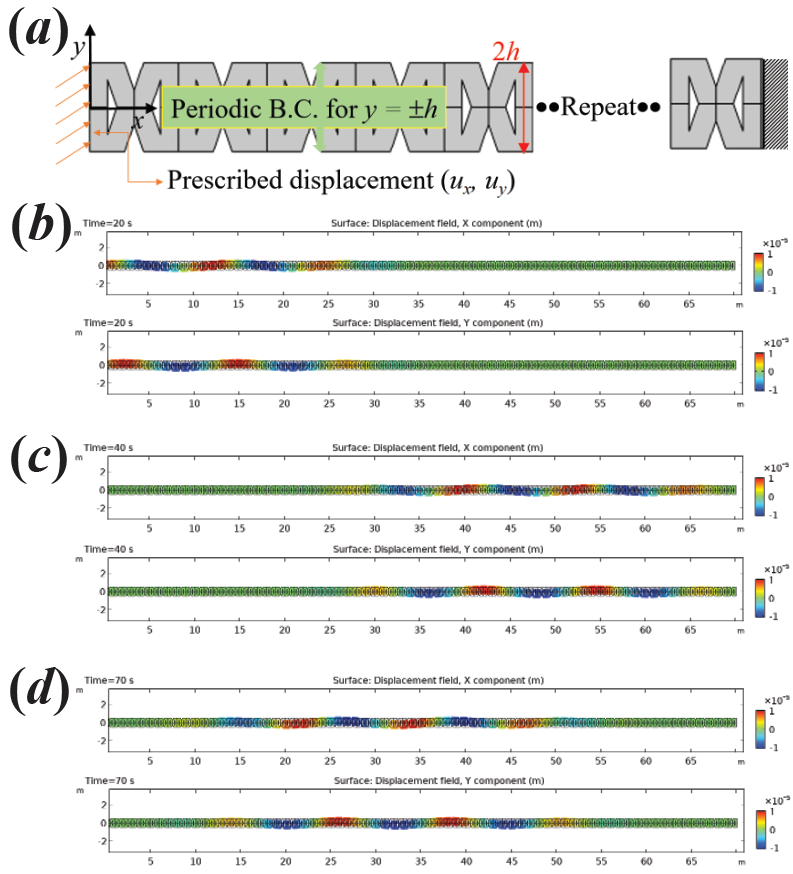}
\caption{\label{fig:SIF4} (a) The schematic of the setup of \textsc{comsol} micro-structured plane model. (b) The displacement fields $u_x$ and $u_y$ at time $t=20$s. (c) The displacement fields $u_x$ and $u_y$ at time $t=40$s. (d) The displacement fields $u_x$ and $u_y$ at time $t=70$s.  }
\end{figure}

\\ 
Fig.\,\ref{fig:SIF3} shows the time-domain finite element simulations of the rolling wave inside 2D anisotropic plane with required elastic constants. This illustrates the satisfaction of the equal-speed criterion and the feasibility for anisotropic material to host the propagating rolling wave.\\
Fig.\,\ref{fig:SIF4} shows the time-domain finite element simulations of the rolling wave inside the designed structured plane. This illustrates the capability of the structure to host the propagating rolling wave. By monitoring specific point, time evolution of the displacements data was extracted and analyzed. It is found the spin property is preserved with the fully-fixed are fully-free boundary conditions, while being flipped with the hybrid fixed-free and hybrid free-fixed boundary conditions.\\
The results elucidate that, in both models, the longitudinal and transverse waves propagate at the same speed. The numerical observations verify that the elastic constants and the structure can host rolling waves. Moreover, the rolling wave spin property can be altered by different boundary conditions. This provides us the potential to use simple edges and surfaces in future applications of rolling waves.

\clearpage
\section{Numerical Procedures}

The 2D and 3D unit cell geometries are designed by finite element calculations in  \textsc{abaqus}  to satisfy the requirements of elastic constants for different anisotropic cases. \\
For 2D plane strain cases, square unit cells are used. We first build one quarter of the unit cell and its mesh. Then, by symmetry operations, the other parts with mesh are generated. This gives us the easiest way to guarantee the one-to-one correspondence between each boundary node-pair, making the application of proper periodic boundary conditions possible. By prescribing unit cell deformation, the effective elastic constants can be obtained by averaging element stresses.\\
Similarly, for 3D cases, we first build the one-eighth structure and then make the symmetry operations. With periodic boundary conditions, we prescribe the unit cell deformation and calculate the average stress components to obtain the effective elastic constants.\\
In addition, 2D time-domain simulations by  \textsc{comsol} are conducted to illustrate the reflections of roll waves from different boundaries. We use two different time-domain models: a) the anisotropic media with elastic constants satisfying the requirements listed in Eq.\,(11a) of the main text; and b) the periodic micro-structured lattice with unit geometry shown in Fig.\,2(a) of the main text. The parameters used in the simulations are set to be at the long-wavelength limit with $\lambda_0/a=4\pi$ ($\lambda_0$ is the wave length, $a$ is the unit cell size). For both models, periodic boundary conditions are applied to the top and bottom boundaries. The displacement boundary conditions with rolling excitation $(u_x,u_y)=({\rm sin}(\omega t), {\rm cos}(\omega t))$ are prescribed at the left edge. The right side as the reflection surface is prescribed with different boundary conditions, i.e., fixed, stress-free, hybrid fixed-free and hybrid free-fixed. \\

\noindent Since numerical procedures employed in this study might be useful in a variety of applications, we make the codes available for free download, advocating for an open-source initiative in the research community:\\
\begin{enumerate}
\item ``ABAQUS-2D.zip" - An Abaqus Python script for the 2D geometry in Fig. 2(a).
\item ``ABAQUS-3D.zip" - An Abaqus Python script for the 3D geometry in Fig. 3(c).
\item ``COMSOL.zip" - A time-domain simulation for results in \ref{fig:SIF4}.
\end{enumerate}
\clearpage

\printbibliography